\documentclass[letterpaper, 12pt]{article}[2000/05/19]

\usepackage[english]{babel}

\usepackage{amsfonts,amsmath,amssymb,amsthm,latexsym,amscd,mathrsfs}
\usepackage{ifthen,cite}
\usepackage[bookmarksnumbered=true]{hyperref}
\hypersetup{pdfpagetransition={Split}}

\newcommand{\evenhead}{Author \ name}
\newcommand{\oddhead}{Article \ name}
\newcommand{\theArticleName}{Article name}

\newcommand{\FirstPageHeading}[1]{\thispagestyle{empty}%
\noindent\raisebox{0pt}[0pt][0pt]{\makebox[\textwidth]{\protect\footnotesize \sf }}\par}

\newcommand{\ArticleName}[1]{\renewcommand{\theArticleName}{#1}\vspace{-2mm}\par\noindent {\LARGE\bf  #1\par}}
\newcommand{\Author}[1]{\vspace{5mm}\par\noindent {\it #1} \par\vspace{2mm}\par}
\newcommand{\Address}[1]{\vspace{2mm}\par\noindent {\it #1} \par}
\newcommand{\Email}[1]{\ifthenelse{\equal{#1}{}}{}{\par\noindent {\rm E-mail: }{\it  #1} \par}}
\newcommand{\URLaddress}[1]{\ifthenelse{\equal{#1}{}}{}{\par\noindent {\rm URL: }{\tt  #1} \par}}
\newcommand{\EmailD}[1]{\ifthenelse{\equal{#1}{}}{}{\par\noindent {$\phantom{\dag}$~\rm E-mail: }{\it  #1} \par}}
\newcommand{\URLaddressD}[1]{\ifthenelse{\equal{#1}{}}{}{\par\noindent {$\phantom{\dag}$~\rm URL: }{\tt  #1} \par}}

\newcommand{\ShortArticleName}[1]{\renewcommand{\oddhead}{#1}}
\newcommand{\AuthorNameForHeading}[1]{\renewcommand{\evenhead}{#1}}

\long\def\@makecaption#1#2{%
  \vskip\abovecaptionskip
  \sbox\@tempboxa{\small \textbf{#1.}\ \ #2}%
  \ifdim \wd\@tempboxa >\hsize
    {\small \textbf{#1.}\ \ #2}\par
  \else
    \global \@minipagefalse
    \hb@xt@\hsize{\hfil\box\@tempboxa\hfil}%
  \fi
  \vskip\belowcaptionskip}

\def\numberwithin#1#2{\@ifundefined{c@#1}{\@nocounterr{#1}}{%
  \@ifundefined{c@#2}{\@nocnterr{#2}}{%
  \@addtoreset{#1}{#2}%
  \toks@\@xp\@xp\@xp{\csname the#1\endcsname}%
  \@xp\xdef\csname the#1\endcsname
    {\@xp\@nx\csname the#2\endcsname
     .\the\toks@}}}}

\newtheorem{theorem}{Theorem}
\newtheorem{lemma}{Lemma}

{\theoremstyle{definition}

}

\newcommand{\abs}[1]{\left\vert#1\right\vert}

\begin {document}

\FirstPageHeading{Shtyk}

\ShortArticleName{The von Neumann Hierarchy for Correlation Operators}

\ArticleName{The von Neumann Hierarchy for Correlation \\ Operators of Quantum Many-Particle Systems}

\Author{V~I Gerasimenko~$^\dag$ V~O Shtyk~$^\ddag$}

\AuthorNameForHeading{V~I Gerasimenko V~O Shtyk}

\Address{$^\dag$ Institute of Mathematics of NAS of Ukraine}

\Address{$^\ddag$ Bogolyubov Institute for Theoretical Physics of NAS of Ukraine}

\EmailD{$^\dag$ gerasym@imath.kiev.ua, $^\ddag$ vshtyk@bitp.kiev.ua}

{\vspace{6mm}\par\noindent\hspace*{8mm}
\parbox{140mm}{\small { $\quad$

The Cauchy problem for the von Neumann hierarchy of nonlinear
equations is investigated. One describes the evolution of all
possible states of quantum many-particle systems by the correlation
operators. A solution of such nonlinear equations is constructed in
the form of an expansion over particle clusters whose evolution is
described by the corresponding order cumulant (semi-invariant) of
evolution operators for the von Neumann equations. For the initial
data from the space of sequences of trace class operators the
existence of a strong and a weak solution of the Cauchy problem is
proved.  We discuss the relationships of this solution both with the
$s$-particle statistical operators, which are solutions of the BBGKY
hierarchy, and with the $s$-particle correlation operators of
quantum systems.
 } }
}

 \makeatletter
\renewcommand{\@evenhead}{
\hspace*{-3pt}\raisebox{-15pt}[\headheight][0pt]{\vbox{\hbox to \textwidth {\thepage \hfil \evenhead}\vskip4pt \hrule}}}
\renewcommand{\@oddhead}{
\hspace*{-3pt}\raisebox{-15pt}[\headheight][0pt]{\vbox{\hbox to \textwidth {\oddhead \hfil \thepage}\vskip4pt\hrule}}}
\renewcommand{\@evenfoot}{}
\renewcommand{\@oddfoot}{}
\makeatother
\newpage
\protect\tableofcontents
\newpage

\section{Introduction}

In the paper we consider the von Neumann hierarchy for correlation
operators that describes the evolution of quantum correlations of
many-particle systems. The necessity to describe the evolution of
correlations arises in many problems of modern statistical
mechanics. Among them we refer to such fundamental problems that are
challenging for mathematics, in  particular, the rigorous derivation
of quantum kinetic equations
\cite{AA,BDGM,BCEP06,BCEP1,BCEP3,BGGM2,ESY,FroL,GP97,Gol,Sp07}, for
example, the kinetic equations describing Bose gases in the
condensate phase \cite{AGT,ESchY1,ESchY2,ESchY3,Schlein}, and the
description of nonequilibrium quantum correlations in ultracold
Fermi and Bose gases \cite{Got}(and references therein). In the
paper we introduce the hierarchy of equations for correlation
operators (the von Neumann hierarchy) that describes the quantum
correlations from the microscopic point of view and shows how such
dynamics is originated in the dynamics of an infinite-particle
system (the BBGKY hierarchy \cite{BogLect,Dop08,Pe95}) and
nonequilibrium fluctuations of macroscopic characteristics of such
systems.

The aim of the work is to formulate the evolution equations
describing correlations in quantum many-particle systems with the
general type of an interaction potential and construct a solution of
the corresponding Cauchy problem, then using the constructed
solution to establish its relationships with the solutions of
hierarchies of the evolution equations of infinitely many quantum
particles.

We outline the structure of the paper and the main results. In
section 2 we introduce preliminary facts about the dynamics of
quantum systems of non-fixed number of particles and deduce the von
Neumann hierarchy for correlation operators which gives the
alternative description of the evolution of states of the
many-particle systems.

In  section 3 we define the cumulants (semi-invariants) of evolution
operators of the von Neumann equation and investigate some of their
typical properties. It turned out that the concept of cumulants of
evolution operators is a basis of the expansions for the solutions
of various evolution equations of quantum systems of particles, in
particular, the von Neumann hierarchy for correlation operators and
the BBGKY hierarchy for infinite-particle systems.

In section 4 the solution of the initial-value problem for the von
Neumann hierarchy is constructed and proved that the solution
generates a group of nonlinear operators of the class $C_{0}$ in the
space of trace class operators. In this space we state an existence
and uniqueness theorem of a strong and a weak solutions for such a
Cauchy problem.

In the last section 5 we discuss the relationships of a solution of
the von Neumann hierarchy for correlation operators both with the
$s$-particle statistical operators, which are solutions of the BBGKY
hierarchy and with the $s$-particle correlation operators of quantum
systems. We also state the general structure of the BBGKY hierarchy
generator of infinite-particle quantum systems.

\section{Evolution equations of quantum many-particle systems}
\subsection{Quantum systems of particles}

We consider a quantum system of a non-fixed
(i.e. arbitrary but finite) number of identical (spinless) particles with unit mass $m=1$ in the space $\mathbb{R}^{\nu},$ $\nu\geq 1$
 (in the terminology of statistical mechanics it is known as a \emph{nonequilibrium grand canonical ensemble}  \cite{CGP97}).

The Hamiltonian of such a system
$H=\bigoplus\limits_{n=0}^{\infty}H_{n}$ is a self-adjoint operator
with domain
                $\mathcal{D}
                (H)=\{\psi=\oplus\psi_{n}\in{\mathcal{F}_{\mathcal{H}}}\mid
                \psi_{n}\in \mathcal{D}(H_{n})\in\mathcal{H}_{n},
               \sum\limits_{n}\|H_{n}\psi_{n}\|^{2}<\infty\}\subset{\mathcal{F}_{\mathcal{H}}},$ where
               $\mathcal{F}_{\mathcal{H}}=\bigoplus\limits_{n=0}^{\infty}\mathcal{H}^{\otimes n}$ is the Fock space over the Hilbert space
                $\mathcal{H}$~ ($\mathcal{H}^{0}=\mathbb{C}$). Assume
                $\mathcal{H}=L^{2}(\mathbb{R}^\nu)$~(coordinate representation) then an element $\psi\in\mathcal{F}_{\mathcal{H}}
                =\bigoplus\limits_{n=0}^{\infty}L^{2}(\mathbb{R}^{\nu n})$ is a sequence of functions
                 $\psi=\big(\psi_0,\psi_{1}(q_1),\ldots,\psi_{n}(q_1,\ldots,q_{n}),\ldots\big)$
such that $\|\psi\|^{2}=|\psi_0|^{2}+\sum_{n=1}^{\infty}\int
dq_1\ldots dq_{n}|\psi_{n}(q_1,\ldots ,q_{n})|^{2}<+\infty.$ On the
subspace of infinitely differentiable functions with compact
supports $\psi_n\in L^{2}_0(\mathbb{R}^{\nu n})\subset
L^{2}(\mathbb{R}^{\nu n})$ $n$-particle Hamiltonian $H_{n}$ acts
according to the formula ($H_{0}=0$)
        \begin{equation}\label{H_Zag}
                H_{n}\psi_n = -\frac{\hbar^{2}}{2}
               \sum\limits_{i=1}^{n}\Delta_{q_i}\psi_n
               +\sum\limits_{k=1}^{n}\sum\limits_{i_{1}<\ldots<i_{k}=1}^{n}\Phi^{(k)}(q_{i_{1}},\ldots,q_{i_{k}})\psi_{n},
        \end{equation}
where  $\Phi^{(k)}$ is a $k$-body interaction potential satisfying Kato conditions
 \cite{Kato} and  $h={2\pi\hbar}$ is a Planck constant.

An arbitrary observable of the many-particle system  $A=(A_{0},A_{1},\ldots,A_{n},\ldots)$ is a sequence of self-adjoint operators $A_{n}$ defined on
the Fock space $\mathcal{F}_{\mathcal{H}}.$ We will consider the observables of the system as the elements of the space
$\mathfrak{L}(\mathcal{F}_\mathcal{H})$ of sequences of bounded operators with an operator norm \cite{DauL_5,Pe95}.

The continuous linear positive functional on the space of observables is defined by the formula
        \begin{equation}\label{averageD}
        \langle A\rangle=\big(e^{\mathfrak{a}}D\big)_{0}^{-1}\sum\limits_{n=0}^{\infty}\frac{1}{n!}
        \mathrm{Tr}_{1,\ldots,n}A_{n}D_{n},
        \end{equation}
where $\mathrm{Tr}_{1,\ldots,n}$ is the partial trace over $1,\ldots,n$ particles,
$\big(e^{\mathfrak{a}}D\big)_{0}=\sum\limits_{n=0}^{\infty}\frac{1}{n!}\mathrm{Tr}_{1,\ldots,n}D_{n}$ is a normalizing factor
(\emph{grand canonical partition function}), for which the notation from section 5 is used.
The sequence $D=(I,D_{1},\ldots,D_{n},\ldots)$ is an infinite sequence of self-adjoint positive density operators $D_{n}$ ($I$ is a unit operator)
 defined on the Fock space
         $\mathcal{F}_{\mathcal{H}}.$ This sequence describes the state of a quantum system of non-fixed number of particles.
The density operators $D_{n},$ $n\geq 1$ (also called
\emph{statistical operators} whose kernels are known as
\emph{density matrices} \cite{BerSh}), defined on the $n$-particle
Hilbert space $\mathcal{H}_{n}=\mathcal{H}^{\otimes
n}=L^{2}(\mathbb{R}^{\nu n})$, we will denote by
$D_{n}(1,\ldots,n)$. For a system of identical particles described
by Maxwell-Boltzmann statistics, one has $
D_{n}(1,\ldots,n)=D_{n}(i_1,\ldots,i_n)$ if
$\{i_{1},\ldots,i_{n}\}\in \{1,\ldots,n\}$.

We consider the states of a system that belong to the space
            $\mathfrak{L}^{1}(\mathcal{F}_\mathcal{H})= \bigoplus\limits_{n=0}^{\infty}
            \mathfrak{L}^{1}(\mathcal{H}_{n})$ of sequences
             $f=\big(I,f_{1},\ldots,f_{n},\ldots\big)$ of trace class operators
            $f_{n}=f_{n}(1,\ldots,n)\in\mathfrak{L}^{1}(\mathcal{H}_{n})$, satisfying the above-mentioned symmetry condition, equipped with the
trace norm
        \begin{eqnarray*}
            \|f\|_{\mathfrak{L}^{1} (\mathcal{F}_\mathcal{H})}=
            \sum\limits_{n=0}^{\infty} \|f_{n}\|_{\mathfrak{L}^{1}(\mathcal{H}_{n})}=
            \sum\limits_{n=0}^{\infty}~\mathrm{Tr}_{1,\ldots,n}|f_{n}(1,\ldots,n)|.
        \end{eqnarray*}
 We will denote by $\mathfrak{L}^{1}_{0}$ the everywhere dense set in $\mathfrak{L}^{1}(\mathcal{F}_\mathcal{H})$ of finite sequences of degenerate
 operators \cite{Kato} with infinitely
differentiable kernels with compact supports. Note that the space
$\mathfrak{L}^{1}(\mathcal{F}_\mathcal{H})$ contains sequences of operators more general than those determining the states of systems of particles.

For the $D\in \mathfrak{L}^{1} (\mathcal{F}_\mathcal{H})$ and $A\in
\mathfrak{L}(\mathcal{F}_\mathcal{H})$ functional (\ref{averageD})
exists. One is interpreted as an average value (expectation value)
of the observable $A$ in the state with density operator  $D$
(nonequilibrium grand canonical ensemble for the Maxwell-Boltzmann
statistics).

We remark that in the case of a system of a fixed number  $N$ of
particles  (\emph{nonequilibrium canonical  ensemble}) the
observables and states are the one-component sequences,
respectively, $A^{(N)}=(0,\ldots,0,A_{N},0,\ldots),$
$D^{(N)}=(0,\ldots,0,D_{N},0,\ldots).$ Therefore, the formula for an
average value reduces to
\begin{equation*}\label{averageD_n}
\langle A^{(N)}\rangle=\big(\mathrm{Tr}_{1,\ldots,N}D_{N}\big)^{-1}\mathrm{Tr}_{1,\ldots,N}A_{N}D_{N}
.
\end{equation*}

\subsection{The von Neumann equation}

The evolution of all possible states $D(t)=(I,D_{1}(t,1),\ldots,D_{n}(t,1,\ldots,n),\ldots)$ is described by the initial-value problem for a sequence
of the von Neumann equations (the quantum Liouville equations) \cite{BogLect,MRS,Pe95}
\begin{eqnarray}
     \label{F-N1}  &&\frac{d}{d t}D(t)=-\mathcal{N}D(t),\\
     \label{F-N12} &&D(t)|_{t=0}=D(0),
\end{eqnarray}
where for $f\in \mathfrak{L}^{1}_{0}(\mathcal{F}_\mathcal{H})\subset\mathcal{D}(\mathcal{N})\subset \mathfrak{L}^{1}(\mathcal{F}_\mathcal{H})$
the von Neumann operator is defined by
\begin{eqnarray}\label{komyt}
   (\mathcal{N}f)_{n}=-\frac{i}{\hbar}\big[f_{n},H_{n}\big]:=
   -\frac{i}{\hbar}\big(f_{n}H_{n}-H_{n}f_{n}\big).
\end{eqnarray}

In the space of sequences of trace-class operators
$\mathfrak{L}^{1}(\mathcal{F}_\mathcal{H})$ for an abstract
initial-value problem (\ref{F-N1})--(\ref{F-N12}) the following
theorem is true.
 \begin{theorem}[\cite{GerSh,Pe95}] The solution of initial-value problem
    (\ref{F-N1})--(\ref{F-N12}) is determined by the formula
    \begin{eqnarray}\label{rozv_fon-N}
        D(t)=\mathcal{U}(-t)D(0)\mathcal{U}^{-1}(-t),
    \end{eqnarray}
     where  ~$\mathcal{U}(-t)=\bigoplus\limits_{n=0}^{\infty}\mathcal{U}_{n}(-t)$  and
    \begin{eqnarray}\label{evol_oper}
        &&\mathcal{U}_{n}(-t)=e^{-{\frac{i}{\hbar}}tH_{n}},\nonumber\\
        &&\mathcal{U}_{n}^{-1}(-t)=e^{{\frac{i}{\hbar}}tH_{n}},
    \end{eqnarray} $\mathcal{U}_{0}(-t)=I$ is a unit operator.

 For  $D(0)\in
    \mathfrak{L}^{1}_{0}(\mathcal{F}_\mathcal{H})\subset
    \mathfrak{L}^{1}(\mathcal{F}_\mathcal{H})$ it is a strong (classical) solution and for arbitrary $D(0) \in
    \mathfrak{L}^{1}(\mathcal{F}_\mathcal{H})$ it is a weak (generalized) solution.
    \end{theorem}

Note that the nature of notations (\ref{evol_oper}) used for unitary
groups $e^{\pm{\frac{i}{\hbar}}tH_{n}}$ is related to the
correspondence principle between quantum and classical systems (for
the later the evolution operator for the Liouville equation for the
density of the distribution function is defined in analogous terms).
It is a consequence of the existence of two approaches to the
description of the evolution of systems based on the description of
the  evolution in framework of observables or states. The evolution
operator generated by solution (\ref{rozv_fon-N}) for
$f\in\mathfrak{L}^{1}(\mathcal{F}_\mathcal{H})$ we will denote by
\begin{equation}\label{groupG}
\mathcal{U}(-t)f\mathcal{U}^{-1}(-t):=\mathcal{G}(-t)f.
\end{equation}

The following properties of the group $\mathcal{G}(-t)$ follow from the properties of groups (\ref{evol_oper}).

In the space $\mathfrak{L}^{1}(\mathcal{F}_\mathcal{H})$
    the mapping $(\ref{groupG}): t\rightarrow
\mathcal{G}(-t)f$
 is an isometric strongly continuous group which preserves positivity and self-adjointness of operators.

For
$f\in\mathfrak{L}_{0}^{1}(\mathcal{F}_\mathcal{H})\subset\mathcal{D}(-\mathcal{N})$
 in the sense of the norm convergence $\mathfrak{L}^{1}(\mathcal{F}_\mathcal{H})$ there exists a limit \cite{Pe95} by which the infinitesimal
 generator: $-\mathcal{N}=\oplus^{\infty}_{n=0}(-\mathcal{N}_{n})$ of the group of evolution operator (\ref{groupG}) is determined
\begin{equation}\label{infOper}
 \lim\limits_{t\rightarrow 0}\frac{1}{t}\big(\mathcal{G}(-t)f-f\big)=-\frac{i}{\hbar}(Hf-fH
):=-\mathcal{N}f,
\end{equation}
 where $H=\oplus^{\infty}_{n=0}H_{n}$ is the Hamiltonian (\ref{H_Zag}) and the operator: $-\frac{i}{\hbar}(Hf-fH)$ is defined in the domain
 $\mathcal{D}(H)\subset\mathcal{F}_\mathcal{H}.$

It should be emphasized that the density operator belonging to the
space $\mathfrak{L}^{1}(\mathcal{F}_\mathcal{H})$ describes only
finitely many-particle systems , i.e. systems for which the average
number of particles in the system is finite. Indeed, for the
observable of the number of particles $N=(0,I,2I,\ldots,nI,\ldots)$
functional (\ref{averageD}) has a form
                \begin{equation}\label{averageND}
                \langle
                N\rangle(t)=\big(e^{\mathfrak{a}}D(0)\big)_{0}^{-1}\sum\limits_{n=0}^{\infty}
                \frac{1}{n!}\mathrm{Tr}_{1,\ldots,n+1}D_{n+1}(t,1,\ldots,n+1),
                \end{equation}
and for
any sequence $D(t)\in\mathfrak{L}^{1}(\mathcal{F}_\mathcal{H})$ we get
\begin{equation*}
|\langle N\rangle(t)| \leq \| D(0)\|_{\mathfrak{L}^{1}(\mathcal{F}_\mathcal{H})} < \infty .
\end{equation*}
\subsection{Derivation of von Neumann hierarchy for correlation operators}
Let us represent the state $D(t)$ of a quantum system  in the form of \emph{cluster expansions}
\cite{BogLect,Rul} over the new operators
$g(t)=(0,g_{1}(t,1),\ldots,g_{n}(t,1,\ldots,n),\ldots)$
                \begin{eqnarray}\label{D_(g)N}
                        &&D_{1}(t,1)=g_{1}(t,1),\nonumber\\
                        &&D_{2}(t,1,2)=g_{2}(t,1,2)+g_{1}(t,1)g_{1}(t,2),\nonumber\\
                        &&\ldots \ldots \ldots\ldots\nonumber\\
                        &&D_{n}(t,Y)=\sum\limits_{\texttt{P}:\,Y=\bigcup\limits_iX_i}
                        \prod_{X_i\subset \texttt{P}}g_{\abs {X_i}}(t,X_i),\,n\geq 1,
                \end{eqnarray} where the following notations are used: $Y\equiv(1,\ldots,n)$, $|Y|=n$ is a number of particles
of the set $Y$,
$\sum\limits_\texttt{P}$ is a sum over all possible partitions $\texttt{P}$ of the set $Y$ into $|\texttt{P}|$ nonempty mutually disjoint
subsets  $X_{i}$.

It is evidently that, in terms of the sequences of operators $g(t),$
the state of the system is described in an equivalent way. The
operators  $g_{n}(t),$ $n\geq 1$ are interpreted as
\emph{correlation operators} of a system of particles.

The evolution of correlation operators is described by the initial-value problem for the von Neumann hierarchy for correlation operators
                \begin{eqnarray}\label{nelNeum1}
                        \frac{d}{dt}g_{n}(t,Y)=-\mathcal{N}_{n}(Y)g_{n}(t,Y)
                   +\sum\limits_{\substack{{\texttt{P}:\,Y=\bigcup\limits
                        X_i},\\|\texttt{P}|>1}}
                                \Big(-\mathcal{N}^{int}(X_{1},\ldots,X_{|\texttt{P}|})\Big)\prod_{X_i\subset \texttt{P}}g_{\abs
                         {X_i}}(t,X_i),
                    \end{eqnarray}
                    \begin{eqnarray}\label{nelNeum2}
                    g_{n}(t,Y)\big|_{t=0}=g_{n}(0,Y),\qquad n\geq1.
                    \end{eqnarray}
 The von Neumann operator $\mathcal{N}_{n}(Y)=\mathcal{N}_{n}$  for the system of particles with Hamiltonian (\ref{H_Zag})
 is defined by formula (\ref{komyt}),
        \begin{eqnarray}\label{oper Nint1}
        \mathcal{N}^{int}(X_{1},\ldots,X_{|\texttt{P}|})=
        \sum\limits_{\substack{{Z_{1}\subset X_{1},}\\Z_{1}\neq \emptyset}}\ldots
        \sum\limits_{\substack{{Z_{|\texttt{P}|}\subset X_{|\texttt{P}|},}\\Z_{|\texttt{P}|}\neq \emptyset}}
        \mathcal{N}_{int}^{\big(\sum\limits_{r=1}^{|\texttt{P}|}|Z_{{r}}|\big)}
        \Big(Z_{{1}},\ldots,Z_{{|\texttt{P}|}}\Big),
        \end{eqnarray}
where $\sum\limits_{Z_{j}\subset X_{j}}$ is a sum over all subsets
$Z_{j}\subset X_{j}$ and for $k=1,\ldots,n$
\begin{eqnarray}\label{oper Nint2}
\mathcal{N}^{(k)}_{int}(1,2,\ldots,k)
=-\frac{i}{\hbar}\big[~\cdot~,\Phi^{(k)}(1,2,\ldots,k)\big].
\end{eqnarray}

The simplest examples of von Neumann hierarchy (\ref{nelNeum1}) are
given by
\begin{eqnarray*}
    &&\frac{d}{dt}g_{1}(t,1)=-\mathcal{N}_{1}(1)g_{1}(t,1),\nonumber\\
    &&\frac{d}{dt}g_{2}(t,1,2)=-\mathcal{N}_{2}(1,2)g_{2}(t,1,2)-\mathcal{N}_{int}^{(2)}(1,2)
    g_{1}(t,1)g_{1}(t,2),\nonumber\\
    &&\frac{d}{dt}g_{3}(t,1,2,3)=-\mathcal{N}_{3}(1,2,3)g_{3}(t,1,2,3)+\nonumber\\
    &&\quad+\big(-\mathcal{N}_{int}^{(2)}(1,2)-\mathcal{N}_{int}^{(2)}(1,3)
    -\mathcal{N}_{int}^{(3)}(1,2,3)\big)
    g_{1}(t,1)g_{2}(t,2,3)+ \nonumber\\
    &&\quad+\big(-\mathcal{N}_{int}^{(2)}(1,2)-\mathcal{N}_{int}^{(2)}(2,3)-
    \mathcal{N}_{int}^{(3)}(1,2,3)\big)
    g_{1}(t,2)g_{2}(t,1,3)+ \nonumber\\
    &&\quad+\big(-\mathcal{N}_{int}^{(2)}(1,3)-\mathcal{N}_{int}^{(2)}(2,3)-
    \mathcal{N}_{int}^{(3)}(1,2,3)\big)
    g_{1}(t,3)g_{2}(t,1,2)+ \nonumber\\
    &&\quad-\mathcal{N}_{int}^{(3)}(1,2,3)g_{1}(t,1)g_{1}(t,2)g_{1}(t,3). \nonumber
\end{eqnarray*}
We note that, in the case of a two-body interaction potential
 $(k=2),$
 the von Neumann hierarchy (\ref{nelNeum1}) is simplified. For example,
 the expression for the operator $g_{3}(t)$ does not contain terms with
   the operator $\mathcal{N}_{int}^{(3)}$. In this case the nonlinear terms in hierarchy (\ref{nelNeum1}) have the form
     \begin{eqnarray}\label{oper NintParn}
        \sum\limits_{\texttt{P}:\,Y=X_{1}\bigcup
        X_2}\,
        \sum\limits_{i_{1}\in X_{1}}
        \sum\limits_{i_{2}\in X_{2}}
        \big(-\mathcal{N}_{int}^{(2)}
        (i_{1},i_{2})\big)g_{|X_{1}|}(t,X_{1})g_{|X_{2}|}(t,X_{2}).
        \end{eqnarray}
For classical systems hierarchy (\ref{nelNeum1}) with nonlinear terms  (\ref{oper NintParn}) is an equivalent form of the corresponding
nonlinear Liouville hierarchy \cite{Sh}
formulated by Green \cite{Gre56}.

The von Neumann hierarchy (\ref{nelNeum1}) can be formally derived
from the
  sequence of (linear) von Neumann equations (\ref{F-N1}) provided that the state of a system is described
  in terms of
  correlation operators defined by cluster expansions (\ref{D_(g)N}).

We remark that, in the case of a system of particles satisfying
Fermi or Bose statistics, cluster expansions (\ref{D_(g)N}) and
hierarchy (\ref{nelNeum1}) have another structure. The analysis of
these cases will be given in a separate paper.

\section{Cluster expansions of evolution operator of von Neumann equation}
\subsection{Cluster expansions}
Let us  expand  the group $\mathcal{G}(-t)$ of evolution operator
(\ref{groupG}) as  follows \emph{(cluster expansions)}:
\begin{eqnarray}\label{groupKlast}
    &\mathcal{G}_{n}(-t,Y)=\sum\limits_{\mathrm{P}:Y =\bigcup\limits_iX_i}
      \prod\limits_{X_i\subset \mathrm{P}}\mathfrak{A}_{|X_i|}(t,X_i),\quad n=|Y| \geq 0,
\end{eqnarray}
where $ \sum\limits_\mathrm{P} $ is the sum over all possible
partitions of the set $ Y\equiv(1,\ldots,n) $ into $|\mathrm{P}|$
nonempty mutually disjoint subsets $ X_i\subset Y.$ The operators
$\mathfrak{A}_{n}(t,Y)$  we refer  to as the \emph{$nth$-order
cumulant (semi-invariant)} of evolution operators (\ref{groupG}).

  The simplest examples of cluster expansions (\ref{groupKlast})
have the form
\begin{eqnarray*}
 &&\mathcal{G}_{1}(-t,1)= \mathfrak{A}_{1}(t,1), \\
 &&\mathcal{G}_{2}(-t,1,2)=
  \mathfrak{A}_{2}(t,1,2)+
  \mathfrak{A}_{1}(t,1) \, \mathfrak{A}_{1}(t,2), \\
 &&\mathcal{G}_{3}(-t,1,2,3)=  \mathfrak{A}_{3}(t,1,2,3)+\mathfrak{A}_{2}(t,1,2) \mathfrak{A}_{1}(t,3)+
               \mathfrak{A}_{2}(t,1,3) \, \mathfrak{A}_{1}(t,2)
                                                         \nonumber \\
  &&\qquad+\mathfrak{A}_{2}(t,2,3) \, \mathfrak{A}_{1}(t,1)+
                 \mathfrak{A}_{1}(t,1)\mathfrak{A}_{1}(t,2)\mathfrak{A}_{1}(t,3).            \nonumber
\end{eqnarray*}
Solving previous relations, one obtains
\begin{eqnarray*}
    &&\mathfrak{A}_{1}(t,1)=\mathcal{G}_{1}(-t,1),\\
    &&\mathfrak{A}_{2}(t,1,2)=\mathcal{G}_{2}(-t,1,2)-\mathcal{G}_{1}(-t,1)\mathcal{G}_{1}(-t,2),
 \\
    &&\mathfrak{A}_{3}(t,1,2,3)=\mathcal{G}_{3}(-t,1,2,3)-\mathcal{G}_{1}(-t,3)\mathcal{G}_{2}(-t,1,2)-\mathcal{G}_{1}(-t,2)\mathcal{G}_{2}(-t,1,3)-\\
    &&\qquad-\mathcal{G}_{1}(-t,1)\mathcal{G}_{2}(-t,2,3)+2!\mathcal{G}_{1}(-t,1)\mathcal{G}_{1}(-t,2)\mathcal{G}_{1}(-t,3).
\end{eqnarray*}

In general case the following lemma is true.

\begin{lemma}
    A solution of  recurrence relations (\ref{groupKlast}) is determined by the
    expansion
    \begin{eqnarray}
    \label{cumulant}
        \mathfrak{A}_{n}(t,Y)
        =\sum\limits_{\mathrm{P}:Y =\bigcup\limits_iX_i}(-1)^{|\mathrm{P}|-1}(|\mathrm{P}|-1)!
        \prod_{X_i\subset \mathrm{P}}\mathcal{G}_{|X_i|}(-t,X_i),\quad\!\!\\
        n = |Y| \geq 1,\quad\!\!\nonumber
    \end{eqnarray}
    where
    $\sum\limits_\mathrm{P} $
    is the sum over all possible partitions of the set
    $Y$
    into
    $|\mathrm{P}|$
    nonempty mutually disjoint subsets
    $ X_i\subset Y.$
\end{lemma}
\begin{proof}
    Let us consider the linear space of sequences
    $f=\big(f_0,f_1(1),\ldots,f_n(1,\ldots,n),\ldots\big)$
    of operators
    $f_n\in\mathfrak{L}^{1}(\mathcal{H}_{n})$    ($f_0$
    is an operator that multiplies a function by an arbitrary number).
    We introduce in this linear space the tensor $\ast$-product \cite{Gen,Rul} 
    \begin{equation}\label{ast}
    (f\ast h)_{|Y|}(Y)=\sum\limits_{Z\subset Y} (f)_{|Z|}(Z)
    \,h_{|Y\backslash Z|} (Y \backslash Z),
    \end{equation} 
where $h=(0,h_{1}(1),\ldots,h_{n}(1,\ldots,n),\ldots)$ is a sequence
    of operators $h_{n}\in\mathfrak{L}^{1}(\mathcal{H}_{n})$
    and $\sum\limits_{Z\subset Y}$
    is the sum over all subsets
    $Z$
    of the set
    $Y\equiv (1,\ldots,n). $

    According to definition (\ref{ast}) for the sequence
    $
     \mathfrak{A}(t)=\big( 0, \mathfrak{A}_1(t,1),
    \mathfrak{A}_{2}(t,1,2),\ldots\linebreak\ldots,\mathfrak{A}_{n}(t,1,\ldots,n),\ldots \big)
    $
    the following equality is true
    $$
    \sum\limits_{\mathrm{P}:Y=\bigcup\limits_iX_i} \
    \prod_{X_i\subset \mathrm{P}}\mathfrak{A}_{(|X_i|)}(t,X_i)=
    \big({\mathbb{E}}
    \mathrm{xp}_{\ast}\mathfrak{A}(t)\big)_{n}(Y),\qquad
        n = |Y| \geq 1,
    $$
    where
    ${\mathbb E}\mathrm{xp}_{\ast}$
    is defined as the $\ast$-exponential mapping, i.e.
    \begin{equation}\label{Exp}
    {\mathbb E} \mathrm{xp}_{\ast}f ={\bf 1}+
    \sum\limits_{n=1}^{\infty} \frac{1}{n!} \underbrace{f \ast \cdots \ast f}_{n},
    \end{equation}
    $f \equiv (f _0,f_1,\ldots,f_n,\ldots)$ and
    ${\bf 1} \equiv ( I, 0 , 0, \ldots )$
    is the unit sequence.

    As a result, we can represent recurrence relations
    (\ref{groupKlast}) in the form
    $$
    {\bf 1}+\tilde{\mathcal{G}}(-t)={\mathbb E}\mathrm{xp}_{\ast} \mathfrak{A}(t),
    $$
    where $\tilde{\mathcal{G}}(-t)=\big(0,\mathcal{G}_{1}(-t,1),\ldots,\mathcal{G}_{n}(-t,1,\ldots,n),\ldots)$ and the elements of the sequence
    $\mathcal{G}(-t)\equiv{\bf 1}+\tilde{\mathcal{G}}(-t)=\big(I,\mathcal{G}_{1}(-t,1),\ldots,\mathcal{G}_{n}(-t,1,\ldots,n),\ldots)$
    are the evolution operators (\ref{groupG}),
    $(\mathcal{G}(-t)f)_{n}(Y)=(\mathcal{U}(-t)f\mathcal{U}^{-1}(-t))_{n}(Y)=\mathcal{U}_{n}(-t,Y)f_{n}(Y)\mathcal{U}_{n}^{-1}(-t,Y)$ for
    $Y=(1,\ldots,n).$

    Similarly, defining the mapping
    ${\mathbb L}\mathrm{n}_{\ast}$
    on the sequences
    $h \equiv (0,h_1,\ldots,h_n,\ldots)$
    as the mapping inverse to
    ${\mathbb E}\mathrm{xp}_{\ast},$
    i.e.
    \begin{equation}\label{Ln}
    {\mathbb L}\mathrm{n}_{\ast} \big( {\bf 1}+h\big)=
    \sum\limits_{n=1}^{\infty} \frac{(-1)^{n-1}}{n} \underbrace{h\ast \cdots \ast h}_{n},
    \end{equation}
    one obtains
    \begin{eqnarray*}
        \sum\limits_{\mathrm{P}:Y =\bigcup\limits_iX_i}(-1)^{|\mathrm{P}|-1}(|\mathrm{P}|-1)!
        \prod_{X_i\subset \mathrm{P}}\mathcal{G}_{|X_i|}(-t,X_i)
        ={\mathbb L}\mathrm{n}_{\ast} \big( {\bf 1}+\tilde{\mathcal{G}}(-t)\big)_{n}(Y),\\
            n = | Y | \geq 1.
    \end{eqnarray*}

 Hence, relation (\ref{cumulant}) can be rewritten as
    \begin{equation*}
        \mathfrak{A}(t)={\mathbb L}\mathrm{n}_{\ast} \big( {\bf 1}
        +\tilde{\mathcal{G}}(-t)\big),
    \end{equation*}
    and, therefore, expression (\ref{cumulant}) is a solution of recurrence relations (\ref{groupKlast}).
\end{proof}
For systems of classical particles cumulants (\ref{cumulant}) were
introduced in \cite{GerR}.

\subsection{Properties of cumulants}
We will now deal with the properties of cumulants (\ref{cumulant}).
As was proved in Lemma 1 the cumulants $\mathfrak{A}_{n}(t),$ $n\geq
1$ of evolution operators (\ref{groupG}) of the von Neumann
equations are solutions of recurrence relations (\ref{groupKlast}),
i.e. cluster expansions of the group of evolution operators
(\ref{groupG}), similar to (\ref{D_(g)N}).

For the quantum system of non-interacting particles for $n\geq 2$ we
have: $\mathfrak{A}_{n}(t)=0.$

Indeed, for a non-interacting Maxwell-Boltzmann gas we have:
$\mathcal{G}_{n}(-t,1,\ldots,n)=\prod_{i=1}^{n}\mathcal{G}_{1}(-t,i)$,
then    \begin{multline*}
        \mathfrak{A}_{n}(t,Y)=
        \sum\limits_{\texttt{P}:\,Y=\bigcup\limits_iX_i}
        (-1)^{| \texttt{P}|-1}(|\texttt{P}|-1)!\prod\limits_{X_{i}\subset \texttt{P}}\prod_{j_i=1}^{|X_{i}|}\mathcal{G}_{1}(-t,j_i)=\\=
        \sum\limits_{k=1}^{n}(-1)^{k-1}\texttt{s}(n,k)(k-1)!\prod_{i=1}^{n}\mathcal{G}_{1}(-t,i)=0.
        \end{multline*}
where the following equality is used:
        \begin{equation}\label{Stirl}
       \sum\limits_{\texttt{P}:\,Y=\bigcup\limits_iX_i}
        (-1)^{| \texttt{P}|-1}(|\texttt{P}|-1)!=
        \sum\limits_{k=1}^{n}(-1)^{k-1}\texttt{s}(n,k)(k-1)!=\delta_{n,1},
   \end{equation} $\texttt{s}(n,k)$ is the Stirling numbers of the second kind and $\delta_{n,1}$ is a Kroneker
   symbol.

In the general case a generator of the $nth$-order cumulant, $n\geq
2,$ is an operator $(-\mathcal{N}^{(n)}_{int})$ defined by an
$n$-body interaction potential (\ref{oper Nint2}).
According to equality (\ref{Stirl}) for the  $nth$-order cumulant, $n\geq2,$ in the sense of point-by-point convergence
of the space $\mathfrak{L}^{1}(\mathcal{H}_{n})$ we have 
\begin{multline}\label{Nint}
       \lim\limits_{t\rightarrow 0}\frac{1}{t}\mathfrak{A}_{n}(t,Y) g_{n}(Y)
       =\sum\limits_{\texttt{P}:\,Y=\bigcup\limits_kZ_k}
        (-1)^{|\texttt{P}|-1}(|\texttt{P}| -1)!\sum\limits_{Z_k\subset \texttt{P}}
        (-\mathcal{N}_{|Z_{k}|}(Z_k))g_{n}(Y)=\\
        =\Big(-\mathcal{N}^{(n)}_{int}(Y)\Big)g_{n}(Y),
\end{multline}
where the operator $\mathcal{N}^{(n)}_{int}$ is given by (\ref{oper Nint2}).

For $n=1$ the generator of the first-order cumulant is given by
        \begin{equation*}
       \lim\limits_{t\rightarrow 0}\frac{1}{t}\big(\mathfrak{A}_{1}(t,Y)-I\big) g_{n}(Y)
       =-\mathcal{N}_{n}(Y)g_{n}(Y),
\end{equation*} where $Y=(1,\ldots,n)$ and  $\big(-\mathcal{N}_{n}(Y)\big)$ is defined by formula (\ref{infOper}).

\section{ Initial-value problem for the von Neumann hierarchy}
\subsection{The formula of a solution}
We consider two approaches to the construction of a solution of the
von Neumann hierarchy (\ref{nelNeum1}).
 Since hierarchy (\ref{nelNeum1}) has the structure of recurrence equations we deduce that the solution
 can be constructed by successive integration of the inhomogeneous von Neumann equations.
 Indeed, for solutions of the first two equations we have
        \begin{eqnarray}
        &&g_{1}(t,1)=\mathcal{G}_{1}(-t,1)g_{1}(0,1),\nonumber\\
        \label{iter2}
        &&g_{2}(t,1,2)=\mathcal{G}_{2}(-t,1,2)g_{2}(0,1,2)+\nonumber\\
        &&+\int\limits_{0}^{t}dt_{1}\mathcal{G}_{2}(-t+t_{1},1,2)
        \big(-\mathcal{N}^{(2)}_{int}(1,2)\big)\mathcal{G}_{1}(-t_{1},1)\mathcal{G}_{1}(-t_{1},2)g_{1}(0,1)g_{1}(0,2).
        \end{eqnarray}

Then let us consider the second term on the right hand side of
(\ref{iter2})
        \begin{eqnarray}\label{iter2kum}
        &&\int\limits_{0}^{t}dt_{1}\mathcal{G}_{2}(-t+t_{1},1,2)
        \big(-\mathcal{N}^{(2)}_{int}(1,2)\big)\mathcal{G}_{1}(-t_{1},1)\mathcal{G}_{1}(-t_{1},2)=\nonumber\\
        &&=-\mathcal{G}_{2}(-t,1,2)\int\limits_{0}^{t}dt_{1}\frac{d}{dt_{1}}\Big(\mathcal{G}_{2}(t_{1},1,2)
        \mathcal{G}_{1}(-t_{1},1)\mathcal{G}_{1}(-t_{1},2)\Big)=\nonumber\\
        &&=\mathcal{G}_{2}(-t,1,2)-\mathcal{G}_{1}(-t,1)\mathcal{G}_{1}(-t,2).
        \end{eqnarray}
The operator
$\mathcal{G}_{2}(-t,1,2)-\mathcal{G}_{1}(-t,1)\mathcal{G}_{1}(-t,2):=\mathfrak{A}_{2}(t,1,2)$
in (\ref{iter2kum}) is the second-order cumulant of evolution
operators (\ref{groupG}). Formula (\ref{iter2kum}) is an analog of
the Duhamel formula, which holds rigorously, for example, for
bounded interaction potential \cite{BanArl}.

Making use of the transformations similar to (\ref{iter2kum}), for
$n>2$ a solution of equations (\ref{nelNeum1}), constructed by the
iterations is presented by expressions (\ref{rozvNF-N}).

The formula for a solution of the von Neumann hierarchy  (\ref{nelNeum1})--(\ref{nelNeum2}) can also be (formally) derived from the solution
$D(t)=(I, D_{1}(t,1),\ldots,D_{n}(t,1,\ldots,n),\ldots)$ of von Neumann equations (\ref{rozv_fon-N}) (Theorem 1) on the bases of cluster expansions
(\ref{D_(g)N}).

Indeed, a solution of recurrence equations  (\ref{D_(g)N}) is
defined by the expressions
        \begin{equation}\label{g_(D)N}
        g_{n}(t,Y)=
        \sum\limits_{\texttt{P}:\,Y=\bigcup\limits_iX_i}(-1)^{\abs {\texttt{P}}-1}(\abs {\texttt{P}} -1)!\,
        \prod_{X_i\subset \texttt{P}}D_{\abs {X_i}}(t,X_i),\,n\geq 1,
        \end{equation}
             where $\sum\limits_\texttt{P}$ is a sum over all possible partitions  $\texttt{P}$
of the set $Y\equiv(1,\ldots,n)$ into $|\texttt{P}|$ nonempty
mutually disjoint subsets $X_{i}$. If we substitute solution
(\ref{rozv_fon-N}) in expressions (\ref{g_(D)N}) and taking into
account cluster expansions (\ref{D_(g)N}) for $t=0$, we derive
        \begin{multline}\label{g(g(0))N}
     g_{n}(t,Y)=\sum\limits_{\texttt{P}:\,Y=\bigcup\limits_iX_i}
     (-1)^{\abs {\texttt{P}}-1}(\abs {\texttt{P}} -1)!
        \prod_{X_i\subset \texttt{P}}\mathcal{G}_{|X_{i}|}(-t,X_{i})\sum\limits_{\texttt{P}_{i}:\,X_{i}=\bigcup\limits_{k_{i}}Z_{k_i}}
                \prod_{Z_{k_i}\subset \texttt{P}_{i}}g_{\abs {Z_{k_i}}}(0,Z_{k_i}).
           \end{multline}
As a result of the regrouping in expression (\ref{g(g(0))N}) the items with similar products of initial operators $\prod_{X_i\subset
\texttt{P}}g_{\abs {X_i}}(0,X_i),$ one obtains
    \begin{equation}\label{rozvNF-N}
     g_{n}(t,Y)=\sum\limits_{\texttt{P}:\,Y=\bigcup\limits_iX_i}
        \mathfrak{A}_{|\texttt{P}|}(t,Y_{\texttt{P}})
        \prod_{X_i\subset \texttt{P}}g_{\abs {X_i}}(0,X_i),\quad n\geq1,
      \end{equation}
    where $Y=(1,\ldots,n)$,
    $Y_{\texttt{P}}\equiv(X_{1},\ldots,X_{|\texttt{P}|})$ is a set whose elements are $|\texttt{P}|$ subsets $X_{i}\subset Y$ of
    partition $\texttt{P}:\,Y=\bigcup\limits_iX_i$ and
     $\sum\limits_{\texttt{P}:\,Y=\bigcup\limits_iX_i}$ is a sum over all possible partitions $\texttt{P}$
of the set $Y$ into $|\texttt{P}|$ nonempty mutually disjoint
subsets. Evolution operators $\mathfrak{A}_{|\texttt{P}|}(t)$ for
every $|\texttt{P}|\geq 1$ in expression (\ref{rozvNF-N}) are
defined by
        \begin{eqnarray} \label{cumulantP}
    \mathfrak{A}_{|\texttt{P}|}(t,Y_{\texttt{P}}):=\sum\limits_{\texttt{P}^{'}:\,Y_{\texttt{P}}=
    \bigcup\limits_kZ_k}
    (-1)^{|\texttt{P}^{'}|-1}({|\texttt{P}^{'}|-1})!
    \prod\limits_{Z_k\subset\texttt{P}^{'}}\mathcal{G}_{|Z_{k}|}(-t,Z_{k}).
    \end{eqnarray}
For $|\texttt{P}|\geq 2$ the  $|\texttt{P}|th$-order cumulants
$\mathfrak{A}_{|\texttt{P}|}(t)$
    of evolution operators (\ref{groupG}) of the von Neumann equations \cite{GerS} have similar structure, in contrast to the first-order
    cumulant.
For example, for $|\texttt{P}|=1$
\begin{eqnarray*}
\mathfrak{A}_{1}(t,1\cup \ldots \cup n)=\mathcal{G}_{n}(-t,1,\ldots,n),
\end{eqnarray*}
for $|\texttt{P}|=2$
\begin{eqnarray*}
\mathfrak{A}_{2}(t,i_{1}\cup\ldots\cup i_{|X_{1}|},i_{|X_{1}|+1}\cup\ldots\cup i_{|Y|})=
\end{eqnarray*}
$$=\mathcal{G}_{|Y|}(-t,1,\ldots,n)-\mathcal{G}_{|X_{1}|}(-t,i_{1},\ldots, i_{|X_{1}|})
\mathcal{G}_{|X_{2}|}(-t,i_{|X_{1}|+1},\ldots, i_{n}),$$ where
$\{i_{1},\ldots, i_{|Y|}\}\in\{1,\ldots,n\}$ and the following
notations are used: the symbol $(i_{1}\cup\ldots\cup
i_{|X_{1}|},i_{|X_{1}|+1}\cup\ldots\cup i_{|Y|})$ denote that the
sets $\{i_{1},\ldots, i_{|X_{1}|}\}$ and $\{i_{|X_{1}|+1},\ldots,
i_{|Y|}\}$ are the connected subsets (clusters, respectively, of
$|X_{1}|$ and $|X_{2}|$ particles) of a partition of the set
$Y=(1,\ldots,|X_{1}|,|X_{1}|+1,\ldots,|Y|)$ into two
 elements.

The simplest examples for solution (\ref{rozvNF-N}) are given by
\begin{eqnarray*}
    &&g_{1}(t,1)=\mathfrak{A}_{1}(t,1)g_{1}(0,1),\\
    &&g_{2}(t,1,2)=\mathfrak{A}_{1}(t,1\cup 2)g_{2}(0,1,2)+\mathfrak{A}_{2}(t,1,2)g_{1}(0,1)
    g_{1}(0,2),\\
    &&g_{3}(t,1,2,3)= \mathfrak{A}_{1}(t,1\cup 2\cup 3)g_{3}(0,1,2,3)+ \mathfrak{A}_{2}(t,2\cup 3,1)g_{1}(0,1)g_{2}(0,2,3)+
    \\    &&\quad\qquad\qquad+\mathfrak{A}_{2}(t,1\cup 3,2)g_{1}(0,2)g_{2}(0,1,3)+
        \mathfrak{A}_{2}(t,1\cup 2,3)g_{1}(0,3)g_{2}(0,1,2)+
    \\    &&\quad\qquad\qquad+\mathfrak{A}_{3}(t,1,2,3)g_{1}(0,1)g_{1}(0,2)g_{1}(0,3).
\end{eqnarray*}

We remark that at the initial instant  $t=0$ solution
(\ref{rozvNF-N}) satisfies initial condition (\ref{nelNeum2}).
Indeed, for $|\texttt{P}|\geq2$ according to definitions
(\ref{evol_oper}) ($\mathcal{U}^{\pm 1}_{n}(0)=I$ is a unit
operator) and in view of equality (\ref{Stirl}), we have
\begin{equation*}\mathfrak{A}_{|\texttt{P}|}(0,Y_{\texttt{P}})=
      \sum\limits_{\texttt{P}^{'}:\,Y_{\texttt{P}}=\bigcup\limits_kZ_k}
       (-1)^{| \texttt{P}^{'}|-1}(|\texttt{P}^{'}|-1)!I=0.
       \end{equation*}
\subsection{Chaos property}
Let us consider the structure of solution  (\ref{rozvNF-N}) for one physically
       motivated example of the initial data; that is to say, if the initial data for Cauchy problem (\ref{nelNeum1})--(\ref{nelNeum2})
       satisfy the chaos property (statistically independent particles) \cite{CGP97}, i.e. the
       sequence of correlation operators is the following one-component sequence
       \begin{eqnarray}\label{posl_g(0)}
      g(0)=(0,g_{1}(0,1),0,0,\ldots).
      \end{eqnarray}
      In fact, in terms of the sequence $D(0)$ this condition means that
      \begin{eqnarray*}
     D(0)=\big(1,D_{1}(0,1), D_{1}(0,1)D_{1}(0,2),\ldots\big).
     \end{eqnarray*}
  Making use of relation (\ref{g_(D)N}) we obtain initial
  condition (\ref{posl_g(0)}) for correlation operators. A more general property, namely, decay of correlations for the classical system of hard spheres,
  was considered in \cite{GerSCor}. \\
      \indent For initial data (\ref{posl_g(0)}) the formula for solution (\ref{rozvNF-N})
      of the initial-value problem (\ref{nelNeum1})--(\ref{nelNeum2})
       is
simplified and is reduced to the following formula:
      \begin{equation}\label{rozvChaosN}
        g_{n}(t,1,\ldots,n)=\mathfrak{A}_{n}(t,1,\ldots,n)
        \prod_{i=1}^{n}g_{1}(0,i),\quad n\geq 1.
              \end{equation}
              It is clear from (\ref{rozvChaosN}) that, if at the initial instant there are no correlations  in the system,
               the correlations generated by the dynamics of a system are completely governed by the cumulants of evolution operators (\ref{groupG}).

In the case of initial data (\ref{posl_g(0)}) solution
(\ref{rozvChaosN}) of the Cauchy problem for the von Neumann
hierarchy (\ref{nelNeum1})--(\ref{nelNeum2}) may be rewritten in
another form. For $n=1$, we have
        \begin{equation*}
        g_{1}(t,1)=\mathfrak{A}_{1}(t,1)g_{1}(0,1).
                \end{equation*}
Then, within the context of the definition of the first-order
cumulant,
       $\mathfrak{A}_{1}(t)$, and the inverse to it evolution operator
$\mathfrak{A}_{1}(-t)$, we express the correlation operators
$g_{n}(t)$, $n\geq2$ in terms of the one-particle correlation
operator $g_{1}(t),$ making use of formula
(\ref{rozvChaosN}).\\
 Finally, for $n\geq2$ formula (\ref{rozvChaosN})  is given by
         \begin{equation*}
        g_{n}(t,1,\ldots,n)=\widehat{\mathfrak{A}}_{n}(t,1,\ldots,n)
        \prod_{i=1}^{n}g_{1}(t,i),\quad n\geq 2,
              \end{equation*}
              where $\widehat{\mathfrak{A}}_{n}(t,1,\ldots,n)$ is the $nth$-order
              cumulant of the \emph{scattering operators}
              \begin{equation*}
      \widehat{\mathcal{G}}_{t}(1,\ldots,n):=
      \mathcal{G}_{n}(-t,1,\ldots,n)\prod_{k=1}^{n}\mathcal{G}_{1}(t,k),
      \quad n\geq1.
           \end{equation*}
           The generator of the scattering operator $\widehat{\mathcal{G}}_{t}(1,\ldots,n)$ is determined by the
           operator:
           $$-\sum\limits_{k=2}^{n}\sum\limits_{i_{1}<\ldots<i_{k}=1}^{n}
           \mathcal{N}_{int}^{(k)}(i_{1},\ldots,i_{k}),$$ where $\mathcal{N}_{int}^{(k)}$ acts according to
           (\ref{oper Nint2}).

\subsection{Properties of a group of nonlinear operators}
On $\mathfrak{L}^{1}(\mathcal{H}_{n})$ solution (\ref{rozvNF-N})
generates  a group of nonlinear operators of the von Neumann
hierarchy. The properties of this group are described in the
following theorem.
\begin{theorem}
     For  $g_{n}\in\mathfrak{L}^{1}(\mathcal{H}_{n}),$ $n\geq1,$
    the mapping
\emph{\begin{equation}\label{groupKum}
       t\rightarrow
\Big(\mathfrak{A}_{t}(g)\Big)_{n}(Y)\equiv \sum\limits_{\texttt{P}:\,Y=\bigcup\limits_iX_i}
        \mathfrak{A}_{|\texttt{P}|}(t,Y_{\texttt{P}})
        \prod_{X_i\subset \texttt{P}}g_{\abs {X_i}}(X_i)
\end{equation}}
\noindent is a group of nonlinear operators of class $C_{0}$. In the
subspace $\mathfrak{L}^{1}_{0}(\mathcal{H}_{n})\subset
    \mathfrak{L}^{1}(\mathcal{H}_{n})$ the infinitesimal generator
    $\mathcal{N}^{nl}(\cdot)$ of group (\ref{groupKum}) is defined by the operator
      \emph{ \begin{eqnarray}\label{Nnl}
        \big(\mathcal{N}^{nl}(g)\big)_{n}(Y):=-\mathcal{N}_{n}(Y)g_{n}(Y)
   +\sum\limits_{\substack{{\texttt{P}:\,Y=\bigcup\limits
        X_i},\\|\texttt{P}|>1}}
                \Big(-\mathcal{N}^{int}(X_{1},\ldots,X_{|\texttt{P}|})\Big)\prod_{X_i\subset \texttt{P}}g_{\abs
         {X_i}}(X_i),
    \end{eqnarray}}
   where the notations are similar to those in (\ref{oper Nint1}).
\end{theorem}

\begin{proof}    Mapping (\ref{groupKum}) is defined for $g_{n}\in\mathfrak{L}^{1}(\mathcal{H}_{n}),$ $n\geq1$
          and the following inequality holds:
\begin{eqnarray}\label{ocinka}
        &&\|\big(\mathfrak{A}_{t}(g)\big)_{n}\|_{\mathfrak{L}^{1}(\mathcal{H}_{n})}\leq n!e^{2n+1}c^{n},
\end{eqnarray}
where $c:=\max\limits_{\texttt{P}:\,Y=\bigcup\limits_iX_i}\|g_{|X_{i}|}(X_{i})\|_{_{\mathfrak{L}^{1}(\mathcal{H}_{|X_{i}|})}}.$

Indeed, inasmuch as  for $g_{n}\in \mathfrak{L}^{1}(\mathcal{H}_{n})$  the equality holds \cite{GerSh}
$$\mathrm{Tr}_{1,\ldots,n}|\mathcal{G}_{n}(-t)g_{n}|=\|g_{n}\|_{\mathfrak{L}^{1}(\mathcal{H}_{n})},$$ we have
     \begin{multline*}
        \|\big(\mathfrak{A}_{t}(g)\big)_{n}\|_{\mathfrak{L}^{1}(\mathcal{H}_{n})}
        \leq
        \sum\limits_{\texttt{P}:\,Y=\bigcup\limits_iX_i}
        \sum\limits_{\texttt{P}^{'}:\,Y_{\texttt{P}}=\bigcup\limits_kZ_k}
     (|\texttt{P}^{'}|-1)!\prod_{X_i\subset \texttt{P}}\|g_{\abs {X_i}}\|_{\mathfrak{L}^{1}(\mathcal{H}_{|X_{i}|})}\leq\\
    \leq \sum\limits_{\texttt{P}:\,Y=\bigcup\limits_iX_i} c^{|\texttt{P}|}
             \sum\limits_{k=1}^{|\texttt{P}|}s(|\texttt{P}|,k)(k-1)!
        \leq \sum\limits_{\texttt{P}:\,Y=\bigcup\limits_iX_i} c^{|\texttt{P}|}
             \sum\limits_{k=1}^{|\texttt{P}|}k^{|\texttt{P}|-1}\leq n!e^{2n+1}c^{n},
     \end{multline*}
 where $\texttt{s}(|\texttt{P}|,k)$ are the Stirling numbers of the second kind. That is,
  $\big(\mathfrak{A}_{t}(g)\big)_{n}\in \mathfrak{L}^{1}(\mathcal{H}_{n})$ for arbitrary
$t\in\mathbb{R}^{1}$ and  $n\geq 1$.\\
 \indent We can now formulate the group property of the one-parametric family of nonlinear operators $\mathfrak{A}_{t}(\cdot)$
  which are defined by (\ref{groupKum}),
 i.e.
       \begin{eqnarray*}
         \mathfrak{A}_{t_{1}}\big(\mathfrak{A}_{t_{2}}(g)\big)=
         \mathfrak{A}_{t_{2}}\big(\mathfrak{A}_{t_{1}}(g)\big)=
         \mathfrak{A}_{t_{1}+t_{2}}(g).
       \end{eqnarray*}
      Indeed, for $g_{n}\in \mathfrak{L}^{1}(\mathcal{H}_{n})$, $n\geq1$ and for any
       $t_{1},\,t_{2}\in \mathbb{R}^{1}$,
       according to (\ref{rozvNF-N}) and (\ref{cumulant}),
       we have
    \begin{multline*}
      \big(\mathfrak{A}_{t_{1}}\big(\mathfrak{A}_{t_{2}}(g)\big)\big)_{n}(Y)
            =\sum\limits_{\texttt{P}:\,Y=\bigcup\limits_iX_i}
                \mathfrak{A}_{|\texttt{P}|}(t_{1},Y_{\texttt{P}})\prod_{X_i\subset \texttt{P}}
                               \sum\limits_{\texttt{P}_{i}:\,X_{i}=\bigcup\limits_{l_{i}}Z_{l_i}}
                \mathfrak{A}_{|\texttt{P}_{i}|}(t_{2},\{X_{i}\}_{\texttt{P}_{i}})
                \prod_{Z_{l_i}\subset \texttt{P}_{i}}g_{\abs
                {Z_{l_i}}}(Z_{l_i})=\\
        =\sum\limits_{\texttt{P}:\,Y=\bigcup\limits_iX_i}
                \sum\limits_{\texttt{P}^{'}:\,Y_{\texttt{P}}=\bigcup\limits_jQ_j}
                (-1)^{|\texttt{P}^{'}|-1}(|\texttt{P}^{'}|-1)!
                \prod\limits_{Q_{j}\subset\texttt{P}^{'}}\mathcal{G}_{|Q_{j}|}(-t_{1},Q_{j})
                     \times\\
        \times\prod_{X_i\subset \texttt{P}}
                \sum\limits_{\texttt{P}_{i}:\,X_{i}=\bigcup\limits_{l_i}Z_{l_i}}
              \sum\limits_{\texttt{P}^{'}_{i}:\,\{X_{i}\}_{\texttt{P}_{i}}=\bigcup\limits_{k_i}R_{k_i}}
                (-1)^{|\texttt{P}^{'}_{i}|-1}(|\texttt{P}^{'}_{i}|-1)!
\prod\limits_{R_{k_i}\subset\texttt{P}^{'}_{i}}\mathcal{G}_{|R_{k_i}|}(-t_{2},R_{k_i})
 \prod_{Z_{l_i}\subset \texttt{P}_{i}}g_{\abs
                    {Z_{l_i}}}(Z_{l_i}),
       \end{multline*} where
$\{X_{i}\}_{\texttt{P}_{i}}\equiv(Z_{1},\ldots,Z_{|\texttt{P}_{i}|})$
is a set whose elements are $|\texttt{P}_{i}|$ subsets
$Z_{l_i}\subset X_{i}$ of the partition
$\texttt{P}_{i}:\,X_{i}=\bigcup\limits_{l_i}Z_{l_i}.$ Having
collected the items at identical products of the initial data
$g_{n}(0)$, $n\geq1$, and taking into account
         the group property of the evolution operators  $\mathcal{U}_{n}^{\pm 1}(-t)$, $n\geq1$ (\ref{evol_oper}), we obtain
       \begin{multline*}
       \big(\mathfrak{A}_{t_{1}}\big(\mathfrak{A}_{t_{2}}(g)\big)\big)_{n}(Y)
      =\sum\limits_{\texttt{P}:\,Y=\bigcup\limits_iX_i}
          \sum\limits_{\texttt{P}^{'}:\,Y_{\texttt{P}}=\bigcup\limits_kQ_k}
     (-1)^{|\texttt{P}^{'}|-1}(|\texttt{P}^{'}|-1)!
        \prod\limits_{Q_{k}\subset\texttt{P}^{'}}\mathcal{G}_{|Q_{k}|}(-t_{1}-t_{2},Q_{k})
        \times
\\        \times\prod_{X_i\subset \texttt{P}}g_{\abs
        {X_i}}(X_i)
      =\sum\limits_{\texttt{P}:\,Y=\bigcup\limits_iX_i}
         \mathfrak{A}_{|\texttt{P}|}(t_{1}+t_{2},Y_{\texttt{P}})
        \prod_{X_i\subset \texttt{P}}g_{\abs{X_i}}(X_i)=
        \big(\mathfrak{A}_{t_{1}+t_{2}}(g)\big)_{n}(Y).
       \end{multline*}
Similarly, we establish
      \begin{eqnarray*}
         \mathfrak{A}_{t_{2}}\big(\mathfrak{A}_{t_{1}}(g)\big)=
         \mathfrak{A}_{t_{1}+t_{2}}(g).
       \end{eqnarray*}

The strong continuity property of the group $\mathfrak{A}_t(g(0))$
over the parameter $t\in \mathbb{R}^{1}$ is a consequence of the
strong continuity of group (\ref{groupG}) of the von Neumann
equations \cite{DauL_5}. Indeed, according to identity (\ref{Stirl})
the following equality holds
   \begin{equation*}
             \sum\limits_{\texttt{P}:\,Y=\bigcup\limits_iX_i}
             \sum\limits_{\texttt{P}^{'}:\,Y_{\texttt{P}}=\bigcup\limits_kZ_k}
        (-1)^{|\texttt{P}^{'}|-1}(|\texttt{P}^{'}|-1)!\prod_{X_i\subset \texttt{P}}g_{\abs {X_i}}(X_i)=
  g_{n}(Y).
   \end{equation*}
Therefore, for $g_{n}\in\mathfrak{L}^{1}_{0}(\mathcal{H}_{n})
\subset \mathfrak{L}^{1}(\mathcal{H}_{n})$, $n\geq 1$, we have
   \begin{multline*}
          \lim_{t\rightarrow 0}\Big\|\sum\limits_{\texttt{P}:\,Y=\bigcup\limits_iX_i}
                                \sum\limits_{\texttt{P}^{'}:\,Y_{\texttt{P}}=\bigcup\limits_kZ_k}
                                    (-1)^{|\texttt{P}^{'}|-1}(|\texttt{P}^{'}|-1)!\prod\limits_{Z_{k}\subset\texttt{P}^{'}}
                                     \mathcal{G}_{|Z_{k}|}(-t,Z_{k})
                        \prod_{X_i\subset \texttt{P}}g_{\abs {X_i}}(X_i)
        -g_{n}(Y)
        \Big\|_{\mathfrak{L}^{1}(\mathcal{H}_{n})}\\
              \leq\sum\limits_{\texttt{P}:\,Y=\bigcup\limits_iX_i}
             \sum\limits_{\texttt{P}^{'}:\,Y_{\texttt{P}}=\bigcup\limits_kZ_k}
        (|\texttt{P}^{'}|-1)!
         \lim_{t\rightarrow 0}\Big\|\prod\limits_{Z_{k}\subset\texttt{P}^{'}}\mathcal{G}_{|Z_{k}|}(-t,Z_{k})
                        \prod_{X_i\subset \texttt{P}}g_{\abs {X_i}}(X_i)
        -\prod_{X_i\subset \texttt{P}}g_{\abs {X_i}}(X_i)\Big\|_{\mathfrak{L}^{1}(\mathcal{H}_{n})}.
   \end{multline*}
In view of the the fact that group $\mathcal{G}_{n}(-t)$
(\ref{groupG}) (Theorem 1) is a strong continuous, i.e. in the sense
of the norm convergence $\mathfrak{L}^{1}(\mathcal{H}_{n})$ there
exists the limit
\begin{eqnarray*}
\lim\limits_{t\rightarrow0}(\mathcal{G}_{n}(-t)g_{n}-g_{n})=0,
 \end{eqnarray*}
which implies that, for mutually disjoint subsets  $X_{i}\subset Y,$
the following equality is also valid:
\begin{eqnarray*}
\lim_{t\rightarrow 0}(\prod\limits_{Z_{k}\subset\texttt{P}^{'}}\mathcal{G}_{|Z_{k}|}(-t,Z_{k})
                        g_{n}
        -g_{n})=0.
 \end{eqnarray*}
For
$g_{n}\in\mathfrak{L}_{0}^{1}(\mathcal{H}_{n})\subset\mathfrak{L}^{1}(\mathcal{H}_{n}),$
we finally have
\begin{equation*}
\lim_{t\rightarrow 0}\|\big(\mathfrak{A}_{t}(g)\big)_{n}-g_{n}\|_{\mathfrak{L}^{1}(\mathcal{H}_{n})}=0.
    \end{equation*}

We will now construct the generator $\mathcal{N}^{nl}(\cdot)$ of
group  (\ref{groupKum}). Taking into account that for
$g_{n}\in\mathfrak{L}_{0}^{1}(\mathcal{H}_{n})\subset\mathcal{D}(\mathcal{N}_{n}^{nl}(\cdot))$
equality (\ref{infOper}) holds,
  let us differentiate the group $\big(\mathfrak{A}_{t}(g)\big)_{n}\psi_{n}$
for all  $\psi_{n}\in \mathcal{D}(H_{n})\subset\mathcal{H}_{n}.$
According to equality (\ref{Nint}) for the $|\texttt{P}|th$-order
cumulant, $|\texttt{P}|\geq 2$, we obtain
\begin{multline*}
       \lim\limits_{t\rightarrow 0}\frac{1}{t}\mathfrak{A}_{|\texttt{P}|}(t,Y_{\texttt{P}}) g_{n}\psi_{n}
       =\sum\limits_{\texttt{P}^{'}:\,Y_{\texttt{P}}=\bigcup\limits_kZ_k}
        (-1)^{|\texttt{P}^{'}|-1}(|\texttt{P}^{'}| -1)!\sum\limits_{Z_k\subset \texttt{P}^{'}}
        (-\mathcal{N}_{|Z_{k}|}(Z_k))g_{n}\psi_{n}=\\
         =\sum\limits_{\substack{{Z_{1}\subset X_{1},}\\Z_{1}\neq \emptyset}}\ldots
        \sum\limits_{\substack{{Z_{|\texttt{P}|}\subset X_{|\texttt{P}|},}\\Z_{|\texttt{P}|}\neq \emptyset}}
        \Big(-\mathcal{N}_{int}^{\big(\sum\limits_{r=1}^{|\texttt{P}|}|Z_{{r}}|\big)}
        \big(Z_{{1}},\ldots,Z_{{|\texttt{P}|}}\big)\Big)g_{n}\psi_{n}
        =\Big(-\mathcal{N}^{int}(Y_{\texttt{P}})\Big)g_{n}\psi_{n},
\end{multline*}
where
    $Y_{\texttt{P}}\equiv(X_{1},\ldots,X_{|\texttt{P}|})$ is a set whose elements are
 $|\texttt{P}|$ subsets $X_{i}\subset Y$ of partition $\texttt{P}:\,Y=\bigcup\limits_iX_i$ and the operator $\mathcal{N}^{(n)}_{int}$ is given by formula (\ref{oper Nint2}).

Therefore, for group  (\ref{groupKum}) we derive
\begin{multline}\label{derivation}
       \lim\limits_{t\rightarrow 0}\frac{1}{t}\Big(\big(\mathfrak{A}_{t}(g)\big)_{n}-g_{n}\Big)\psi_{n}       =\lim\limits_{t\rightarrow 0}\frac{1}{t}\Big(\sum\limits_{\texttt{P}:\,Y=\bigcup\limits_iX_i}
                \mathfrak{A}_{|\texttt{P}|}(t,Y_{\texttt{P}})
                \prod_{X_i\subset \texttt{P}}g_{\abs {X_i}}(X_i)-g_{n}(Y)\Big)\psi_{n}=\\
            =\lim\limits_{t\rightarrow 0}\frac{1}{t}\big(\mathfrak{A}_{1}(t,Y)g_{n}-g_{n}\big)\psi_{n}
            +  \sum\limits_{\substack{{\texttt{P}:\,Y=\bigcup\limits_i
        X_i},\\|\texttt{P}|>1}}\lim\limits_{t\rightarrow 0}\frac{1}{t}\mathfrak{A}_{|\texttt{P}|}(t,Y_{\texttt{P}})
                \prod_{X_i\subset \texttt{P}}g_{\abs {X_i}}(X_i)\psi_{n}=\\
        =\big(-\mathcal{N}_{n}g_{n}\big)(Y)\psi_{n}
        +\sum\limits_{\substack{{\texttt{P}:\,Y=\bigcup\limits_i
        X_i},\\|\texttt{P}|>1}}
                \Big(-\mathcal{N}^{int}(Y_{\texttt{P}})\Big)\prod_{X_i\subset \texttt{P}}g_{\abs
         {X_i}}(X_i)\psi_{n}.
\end{multline}

Then in view of equality (\ref{derivation}) and the proof of the
theorem 2
 for $g_{n}\in\mathfrak{L}_{0}^{1}(\mathcal{H}_{n})\subset\mathcal{D}
 (\mathcal{N}_{n}^{nl}(\cdot))\subset\mathfrak{L}^{1}(\mathcal{H}_{n}),\, n\geq1,$ in the sense of the norm convergence
 in $\mathfrak{L}^{1}(\mathcal{H}_{n}),$
  we finally have
\begin{eqnarray*}
 \lim_{t\rightarrow 0}\Big\| \frac{1}{t}\big(\big(\mathfrak{A}_{t}(g)\big)_{n}-
 g_{n}\big)
 -\big(\mathcal{N}^{nl}(g)\big)_{n}\Big\|_{\mathfrak{L}^{1}(\mathcal{H}_{n})}=0,
\end{eqnarray*}
where $\mathcal{N}^{nl}(\cdot)$ is given by formula (\ref{Nnl}).

\end{proof}

\subsection{The uniqueness and existence theorem} For abstract initial-value problem  (\ref{nelNeum1})--(\ref{nelNeum2})
in the space $\mathfrak{L}^{1}(\mathcal{H}_{n})$  of trace class operators the following theorem holds.
\begin{theorem} The solution of initial-value problem
    (\ref{nelNeum1})--(\ref{nelNeum2}) for the von Neumann hierarchy  (\ref{nelNeum1}) is determined by formula (\ref{rozvNF-N}).
     For $g_{n}(0)\in
    \mathfrak{L}^{1}_{0}(\mathcal{H}_{n})\subset
    \mathfrak{L}^{1}(\mathcal{H}_{n})$ it is a strong (classical) solution and for arbitrary initial data $g_{n}(0) \in
    \mathfrak{L}^{1}(\mathcal{H}_{n})$ it is a weak (generalized) solution.
\end{theorem}
\begin{proof} According to theorem 2 for initial data
$g_{n}(0)\in\mathfrak{L}_{0}^{1}(\mathcal{H}_{n})\subset\mathfrak{L}^{1}(\mathcal{H}_{n}),\,
n\geq1,$ sequence (\ref{rozvNF-N}) is a strong solution of
initial-value problem
 (\ref{nelNeum1})-(\ref{nelNeum2}).

Let us show that in the general case $g(0)\in\mathfrak{L}^{1}(\mathcal{F}_\mathcal{H})$
 expansions (\ref{rozvNF-N}) give a weak solution of
the initial-value problem for the von Neumann hierarchy
(\ref{nelNeum1}). Consider the functional
\begin{eqnarray}\label{func-g}
&&\big(\varphi_{n},g_{n}(t)\big):=
    \mathrm{Tr}_{1,\ldots,n}\,\varphi_{n}g_{n}(t),
\end{eqnarray}
where $\varphi_{n}\in \mathfrak{L}_{0}(\mathcal{H}_{n})$ are
degenerate bounded operators with infinitely times differentiable
kernels with compact supports. The operator $g_{n}(t)$ is defined by
(\ref{rozvNF-N}) for the arbitrary initial data
$g_{k}(0)\in\mathfrak{L}^{1}(\mathcal{H}_{k}),$ $k=1,\ldots,n.$
According to estimate (\ref{ocinka}) for $g_{n}\in
\mathfrak{L}^{1}(\mathcal{H}_{n})$ and $\varphi_{n}\in
\mathfrak{L}_{0}(\mathcal{H}_{n})$ functional (\ref{func-g}) exists.

We transform  functional (\ref{func-g}) as follows
\begin{eqnarray}\label{funk-gN}
    &&\big(\varphi_{n},g_{n}(t)\big)=\sum\limits_{\texttt{P}:Y=\bigcup\limits_iX_i}\Big(\varphi_{n},
                \mathfrak{A}_{|\texttt{P}|}(t,Y_{\texttt{P}})
                \prod_{X_i\subset \texttt{P}}g_{\abs
                {X_i}}(0,X_i)\Big)=\nonumber\\
                &&=\sum\limits_{\texttt{P}:\,Y=\bigcup\limits_iX_i}
                \sum\limits_{\texttt{P}^{'}:\,Y_{\texttt{P}}=\bigcup\limits_{k}Z_{k}}
                (-1)^{|\texttt{P}^{'}|-1}(|\texttt{P}^{'}|-1)!\Big(\prod\limits_{Z_{k}\subset\texttt{P}^{'}}
                \mathcal{G}_{|Z_{k}|}(t;Z_{k})\varphi_{n},
                \prod\limits_{X_{i}\subset\texttt{P}}
                g_{\abs {X_{i}}}(0,X_{i})\Big),
\end{eqnarray}
where the group of operators
 $\mathcal{G}_{n}(t)$ is adjoint to the group $\mathcal{G}_{n}(-t)$ in the sense of functional (\ref{func-g}).

For $g_{n}(0)\in \mathfrak{L}^{1}(\mathcal{H}_{n})$ and
$\varphi_{n}\in \mathfrak{L}_{0}(\mathcal{H}_{n})$ within the
context of the theorem  2
 we have
\begin{eqnarray}
\lim\limits_{t\rightarrow0}\Big(\big(\frac{1}{t}(\mathcal{G}_{n}(t)\varphi_{n}-\varphi_{n}),g_{n}(0)\big)
-\big(\mathcal{N}_{n}\varphi_{n},g_{n}(0)\big)\Big)=0,\nonumber
\end{eqnarray}
and, therefore, it holds that
\begin{eqnarray*}
    &&\lim\limits_{t\rightarrow0}\Big(\frac{1}{t}\big(\prod\limits_{Z_{k}\subset\texttt{P}^{'}}
                \mathcal{G}_{|Z_{k}|}(t;Z_{k})\varphi_{n}-\varphi_{n}\big),
                \prod\limits_{X_{i}\subset\texttt{P}}
                g_{\abs {X_{i}}}(0,X_{i})\Big)=\\
                &&=\Big(\sum\limits_{Z_{l}\subset\texttt{P}^{'}}
                \mathcal{N}_{|Z_{l}|}(Z_{l})\varphi_{n},
                \prod\limits_{Z_{k}\subset\texttt{P}^{'}}
                \mathcal{G}_{|Z_{k}|}(-t;Z_{k})\prod\limits_{X_{i}\subset\texttt{P}}
                g_{\abs {X_{i}}}(0,X_{i})\Big).
\end{eqnarray*}
Then, for representation (\ref{funk-gN}) of functional (\ref{func-g}), one obtains
\begin{eqnarray*}
    &&\frac{d}{dt}\big(\varphi_{n},g_{n}(t)\big)=\Big(\mathcal{N}_{n}\varphi_{n},g_{n}(t)\Big)
                +\sum\limits_{\substack{{\texttt{P}:\,Y=\bigcup\limits_i
                X_i},\\|\texttt{P}|>1}}
                \sum\limits_{\texttt{P}^{'}:\,Y_{\texttt{P}}=\bigcup\limits_{k}Z_{k}}
                (-1)^{|\texttt{P}^{'}|-1}(|\texttt{P}^{'}|-1)!\times\nonumber\\
                &&\times\Big(\sum\limits_{Z_{l}\subset\texttt{P}^{'}}
                \mathcal{N}_{|Z_{l}|}(Z_{l})\varphi_{n},
                \prod\limits_{Z_{k}\subset\texttt{P}^{'}}
                \mathcal{G}_{|Z_{k}|}(-t;Z_{k})\prod\limits_{X_{i}\subset\texttt{P}}
                g_{\abs {X_{i}}}(0,X_{i})\Big)=\\
                &&=\Big(\mathcal{N}_{n}\varphi_{n},g_{n}(t)\Big)+\sum\limits_{\substack{{\texttt{P}:\,Y=\bigcup\limits_i
                X_i},\\|\texttt{P}|>1}}
                \Big(\mathcal{N}^{int}(X_{1},\ldots,X_{|\texttt{P}|})\varphi_{n},\prod_{X_i\subset \texttt{P}}g_{\abs
                {X_i}}(t,X_i)\Big),
\end{eqnarray*}
where the operator
$\mathcal{N}^{int}(X_{1},\ldots,X_{|\texttt{P}|})$ is defined by
(\ref{oper Nint1}) and (\ref{oper Nint2}).

For functional (\ref{func-g})
 we finally have
\begin{multline}\label{d_funk-gN}
    \frac{d}{dt}\big(\varphi_{n},g_{n}(t)\big)=\\
    =\Big(\big(\mathcal{N}_{n}\varphi_{n}\big)(Y),g_{n}(t,Y)\Big)
                +\sum\limits_{\substack{{\texttt{P}:\,Y=\bigcup\limits_i
        X_i},\\|\texttt{P}|>1}}
                \Big(\mathcal{N}^{int}(X_{1},\ldots,X_{|\texttt{P}|})\varphi_{n},\prod_{X_i\subset \texttt{P}}g_{\abs
         {X_i}}(t,X_i)\Big).
         \end{multline}
Equality (\ref{d_funk-gN}) means that for arbitrary initial data
$g_{n}(0)\in\mathfrak{L}^{1}(\mathcal{H}_{n}),$ $n\geq 1,$ the weak
solution of the initial-value problem of the von Neumann hierarchy
(\ref{nelNeum1})--(\ref{nelNeum2}) is determined by formula
(\ref{rozvNF-N}).
\end{proof}

\section{BBGKY hierarchy}
\subsection{Nonequilibrium grand canonical ensemble}
As we have seen above the two equivalent approaches to the the
description of the state evolution of quantum many-particle systems
were formulated, namely, both on the basis of von Neumann equations
(\ref{F-N1}) for the statistical operators $D(t)$ and of the von
Neumann hierarchy (\ref{nelNeum1}) for the correlation operators
$g(t)$. For the system of a finite average number of particles there
exists another possibility to describe the evolution of states,
namely, by sequences of $s$-particle statistical operators that
satisfy the BBGKY hierarchy \cite{BogLect}.

Traditionally such a hierarchy is deduced on the basis of solutions
of the von Neumann equations (the nonequilibrium grand canonical
ensemble \cite{CGP97,GerS,Pe95} or the canonical ensemble
\cite{AA,BogLect,ESY}) in the space of sequences of trace class
operators.

The sequence $F(t)=(I,F_{1}(t,1),\ldots,F_{s}(t,1,\ldots,s),\ldots)$
of $s$-particle statistical operators $F_{s}(t,1,\ldots,s),$ $s\geq
1$ can be defined in the framework of the sequence of operators
  $D(t)=(I, D_{1}(t,1),\ldots,D_{n}(t,1,\ldots,n),\ldots)$ (the operator $D_{n}(t)$ being regarded as density operator (\ref{rozv_fon-N})
  of the
  $n$-particle system) by expressions \cite{Pe95}
\begin{equation}\label{F(D)}
F_{s}(t,1,\ldots,s)=\big(e^\mathfrak{a}D(0)\big)^{-1}_{0}
      \sum\limits_{n=0}^{\infty}\frac{1}{n!}\mathrm{Tr}_{s+1,\ldots,s+n}D_{s+n}
      (t,1,\ldots,s+n), \quad s\geq1,
\end{equation}
where $\big(e^\mathfrak{a}D(0)\big)_{0}=1+\sum\limits_{n=1}^{\infty}
\frac{1}{n!}\mathrm{Tr}_{1,\ldots,n}D_{n}(0,1,\ldots,n)$ is a
partition function (see definition of functional (\ref{averageD})).
For $D(0)\in\mathfrak{L}^{1}(\mathcal{F}_\mathcal{H})$ series
(\ref{F(D)}) converges.

If we describe the states of a quantum system of particles in the
framework of correlation operators $g(t)$ the $s$-particle
statistical operators, that are solutions of the BBGKY hierarchy,
are defined by the expansion
\begin{equation}\label{F(g)}
F_{s}(t,1,\ldots,s)=\sum\limits_{n=0}^{\infty}\frac{1}{n!}\mathrm{Tr}_{s+1,\ldots,s+n}g_{1+n}(t,1\cup\ldots\cup s,s+1,\ldots,s+n),\quad
s\geq1,
\end{equation}
where $g_{1+n}(t,1\cup\ldots\cup s,s+1,\ldots,s+n),$ $n\geq 0,$ are
correlation operators (\ref{rozvNF-N}) that satisfy the von Neumann
hierarchy (\ref{nelNeum1}) for a system consisting both from
particles and the cluster of $s$ particles ($1\cup\ldots\cup s$ is a
notation as to formula (\ref{cumulantP})).

Expansion (\ref{F(g)}) can be derived from (\ref{F(D)}) as a result of the following representation for right hand side of expansion (\ref{F(D)})
\begin{multline}\label{F(D)cl}
F_{s}(t,1,\ldots,s)=\\
=\sum\limits_{n=0}^{\infty}\frac{1}{n!}\mathrm{Tr}_{s+1,\ldots,s+n} \sum\limits_{\texttt{P}:\{1\cup\ldots\cup s,s+1,\ldots,s+n\}=\bigcup\limits_iX_i}(-1)^{\abs {\texttt{P}}-1}(\abs {\texttt{P}} -1)!\,
        \prod_{X_i\subset \texttt{P}}D_{\abs {X_i}}(t,X_i),\,s\geq 1,
\end{multline} where $\sum\limits_{\texttt{P}}$ is a sum over all possible partitions $\texttt{P}$
of the set $\{1\cup\ldots\cup s,s+1,\ldots,s+n\}$  into $|\texttt{P}|$ nonempty mutually disjoint subsets $X_{i}$.
If $D(t)\in \mathfrak{L}^{1}(\mathcal{F}_\mathcal{H})$ series (\ref{F(D)cl}) converges.

We remark that expansion (\ref{F(D)}) can be defined for more general class of operators then from the space $\mathfrak{L}^{1}(\mathcal{F}_\mathcal{H}).$

To prove representation (\ref{F(D)cl}) we will first introduce some
necessary facts. We will use the notations of lemma 1.
    According to the definition of $\ast$-product (\ref{ast}) for the sequence
    $
     f=\big(f_{0}, f_1(1),f_{2}(1,2),\ldots,f_{n}(1,\ldots,n),\ldots \big)$ of elements $f_{n}\in\mathfrak{L}^{1}(\mathcal{H}_{n})$
    the mapping     ${\mathbb E}\mathrm{xp}_{\ast}$     is defined by series (\ref{Exp}).
The mapping $\mathbb{L}\mathrm{n}_{\ast}$ that inverse to ${\mathbb E}\mathrm{xp}_{\ast}$ is defined by series (\ref{Ln}).

      For $f=(f_{0},f_{1},\ldots,f_{n},\ldots),$  $f_{n}\in\mathfrak{L}^{1}(\mathcal{H}_{n})$ we define the mapping $\mathfrak{d}_{1}:f\rightarrow
\mathfrak{d}_{1}f$  by
      \begin{equation*}
      (\mathfrak{d}_{1}f)_{n}(1,\ldots,n):=f_{n+1}(1,\ldots,n,n+1),\,\,\,n\geq0,
      \end{equation*}
and for arbitrary set $Y=(1,\ldots,s)$  we define the linear mapping
$\mathfrak{d}_{1}\ldots \mathfrak{d}_{s}:f\rightarrow
\mathfrak{d}_{1}\ldots \mathfrak{d}_{s}f$ by
      \begin{equation}\label{oper_d}
     (\mathfrak{d}_{1}\ldots \mathfrak{d}_{s}f)_{n}:=f_{s+n}(1,\ldots,s+n).
      \end{equation}
     We  note that for sequences $f^{1}=(f^{1}_{0},f^{1}_{1},\ldots,f^{1}_{n},\ldots),$
     $f^{1}_{n}\in\mathfrak{L}^{1}(\mathcal{H}_{n})$ and
     $f^{2}=\linebreak(f^{2}_{0},f^{2}_{1},\ldots,f^{2}_{n},\ldots),$
      $f^{2}_{n}\in\mathfrak{L}^{1}(\mathcal{H}_{n})$  the following identity holds
      \cite{Rul}
      \begin{equation*}
      \mathfrak{d}_{1}(f^{1}\ast f^{2})=\mathfrak{d}_{1}f^{1}\ast f^{2}+f^{1}\ast
      \mathfrak{d}_{1}f^{2}.
      \end{equation*}
Further, since $Y_{\mathrm{P}}=(X_1,\ldots,X_{|\mathrm{P}|})$ then
$Y_{1}=(1\cup\ldots\cup s)$ (cluster of $s$ particles) is  one
element ($|Y_{1}|=1$) of the partition $\mathrm{P}$
($|\mathrm{P}|=1$), we introduce the mapping
$\mathfrak{d}_{Y_{1}}:f\rightarrow \mathfrak{d}_{Y_{1}}f$,  as
follows
      \begin{equation}\label{oper_dKlaster}
      (\mathfrak{d}_{Y_{1}}f)_{n}(1,\ldots,n)=f_{n+1}(Y_{1},1,\ldots,n,),\,\,\,n\geq0.
      \end{equation}
Then for arbitrary $f=(f_{0},f_{1},\ldots,f_{n},\ldots),$
$f_{n}\in\mathfrak{L}^{1}(\mathcal{H}_{n})$ according to definition
(\ref{Exp}) of the mapping  $\mathbb{E}\mathrm{xp}_{\ast}$ the
following equality holds
      \begin{equation}\label{d_gamma}
      \mathfrak{d}_{1}(\mathbb{E}\mathrm{xp}_{\ast} f)=\mathfrak{d}_{1}f\ast\mathbb{E}\mathrm{xp}_{\ast} f.
      \end{equation}
      For $f_{n}\in\mathfrak{L}^{1}(\mathcal{H}_{n})$ an analog of the annihilation operator is defined
      \begin{equation}\label{a}
      (\mathfrak{a}f)_{s}(1,\ldots,s):=
      \mathrm{Tr}_{s+1}f_{s+1}(1,\ldots,s,s+1),
      \end{equation}
      and, therefore
      \begin{equation*}
      (e^{\mathfrak{a}}f)_{s}(1,\ldots,s)=\sum\limits_{n=0}^{\infty}\frac{1}{n!}\mathrm{Tr}_{s+1,\ldots,{s+n}}
      f_{s+n}(1,\ldots,s,s+1,\ldots,s+n),\quad s\geq0.
      \end{equation*}
Using previous definitions and (\ref{ast}) for sequences
$f^{1}=(f^{1}_{0},f^{1}_{1},\ldots,f^{1}_{n},\ldots),$$f^{1}_{n}\in\mathfrak{L}^{1}(\mathcal{H}_{n})$
and $f^{2}=(f^{2}_{0},f^{2}_{1},\ldots,f^{2}_{n},\ldots),$
$f^{2}_{n}\in\mathfrak{L}^{1}(\mathcal{H}_{n})$ we have \cite{Rul}
      \begin{equation}\label{efg}
      (e^{\mathfrak{a}}(f^{1}\ast f^{2}))_{0}=
      (e^{\mathfrak{a}}f^{1})_{0}(e^{\mathfrak{a}}f^{2})_{0}.
      \end{equation}
From equality (\ref{efg}) we deduce that expressions (\ref{F(D)}) and (\ref{F(g)}) can, respectively, be rewritten as
\begin{equation*}
F_{s}(t,Y)=\big(e^\mathfrak{a}D(0)\big)^{-1}_{0}\big(e^\mathfrak{a}\mathfrak{d}_{Y_{1}}D(t)\big)_{0}
\end{equation*}
and
\begin{eqnarray*}
F_{s}(t,Y)=\big(e^\mathfrak{a}\mathfrak{d}_{Y_{1}}g(t)\big)_{0},
\end{eqnarray*}
where $Y=(1,\ldots,s).$

Hence, in view of equalities  (\ref{D_(g)N}), (\ref{Ln}) to derive
expressions (\ref{F(g)}) we have to prove the following lemma.
\begin{lemma}
For $f=(f_{0},f_{1},\ldots,f_{n},\ldots)$ and $f_{n}\in\mathfrak{L}^{1}(\mathcal{H}_{n})$  the following identity holds
      \begin{eqnarray}\label{lema2}
      (e^\mathfrak{a}\mathbb{E}\mathrm{xp}_{\ast}f)^{-1}_{0}\big(e^\mathfrak{a}\mathfrak{d}_{Y_{1}}(\mathbb{E}\mathrm{xp}_{\ast}f)\big)_{0}
      =
      (e^\mathfrak{a}\mathfrak{d}_{Y_{1}}f)_{0}.
      \end{eqnarray}
\end{lemma}
\begin{proof}
Indeed, using equalities (\ref{oper_dKlaster}), (\ref{d_gamma}) and
(\ref{efg}), we obtain
      \begin{eqnarray}
      &&(e^\mathfrak{a}\mathbb{E}\mathrm{xp}_{\ast} f)^{-1}_{0}\big(e^\mathfrak{a}\mathfrak{d}_{Y_{1}}(\mathbb{E}\mathrm{xp}_{\ast} f)\big)_{0}=(e^\mathfrak{a}\mathbb{E}\mathrm{xp}_{\ast} f)^{-1}_{0}\big(e^\mathfrak{a}(\mathbb{E}\mathrm{xp}_{\ast} f\ast
      \mathfrak{d}_{Y_{1}}f)\big)_{0}=\nonumber\\
      &&\qquad=(e^\mathfrak{a}\mathbb{E}\mathrm{xp}_{\ast} f)^{-1}_{0}(e^\mathfrak{a}\mathbb{E}\mathrm{xp}_{\ast} f)_{0}
      \big(e^\mathfrak{a}\mathfrak{d}_{Y_{1}}f\big)_{0}=\big(e^\mathfrak{a}\mathfrak{d}_{Y_{1}}f\big)_{0}.\nonumber
      \end{eqnarray}

\end{proof}

\subsection{On a solution of the BBGKY hierarchy}
In this subsection we will turn to the solution of the initial-value problem of the BBGKY hierarchy.
In the space $\mathfrak{L}^{1}_{\alpha}(\mathcal{F}_\mathcal{H})=\bigoplus\limits_{n=0}^{\infty}\alpha^{n}\mathfrak{L}^{1}(\mathcal{H}_{n})$, where $\alpha>1$ is a real number,
 we consider the following initial-value problem  for
the BBGKY hierarchy for quantum systems of particles obeying
Maxwell-Boltzmann statistics
\begin{eqnarray} \label{1}
        &&\frac{d}{dt}F_{s}(t)=-\mathcal{N}_{s}F_{s}(t)+\nonumber\\
        &&\qquad+\sum\limits_{n=1}^{\infty}\frac{1}{n!}
        \texttt{Tr}_{s+1,\ldots,s+n}\sum\limits_{\substack{{Z\subset Y,}\\Z\neq\emptyset}}
        \Big(-\mathcal{N}_{int}^{(|Z|+n)}(Z,s+1,\ldots,s+n)\Big)
        F_{s+n}(t),\\
        \label{2}
        &&F_{s}(t)\mid_{t=0}=F_{s}(0),\quad s\geq 1,
\end{eqnarray}
where $Y=(1,\ldots,s)$ and  the operator $\mathcal{N}^{(n)}_{int}$
is defined on
$\mathfrak{L}^{1}_{\alpha,0}(\mathcal{F}_\mathcal{H})\subset
\mathfrak{L}^{1}_{\alpha}(\mathcal{F}_\mathcal{H})$ by formula
(\ref{oper Nint2}). The equation from hierarchy (\ref{1})  for $s=1$
has the following transparent form:
                \begin{equation*}
                \frac{d}{dt}F_{1}(t)=-\mathcal{N}_{1}F_{1}(t)
                +\sum\limits_{n=1}^{\infty}\frac{1}{n!}
                \texttt{Tr}_{2,\ldots,n+1}\Big(-\mathcal{N}_{int}^{(n+1)}(1,2,\ldots,n+1)\Big)
                F_{n+1}(t).
                \end{equation*}
In the space $\mathfrak{L}^{1}_{\alpha}(\mathcal{F}_\mathcal{H})$ there is an equivalent representation for the generator of the
BBGKY hierarchy:
$$e^{\mathfrak{a}}(-\mathcal{N})e^{-\mathfrak{a}}=-\mathcal{N}+\sum\limits_{n=1}^{\infty}\frac{1}{n!}[\ldots[\mathcal{N},\underbrace{\mathfrak{a}],\ldots,\mathfrak{a}]}_{n-times}$$
that follows from (\ref{F(D)}). For a two-body interaction potential
one is reduced to the form \cite{GerSh}:
$-\mathcal{N}+\big[\mathcal{N},\mathfrak{a}\big].$

We remark that in terms of the $s$-particle density matrix (marginal
distribution)
$F_{s}(t,q_{1},\ldots,q_{s};q^{'}_{1},\ldots,q^{'}_{s})$ that are
kernels of the $s$-particle density operators $F_{s}(t),$ for a
two-body interaction potential (see (\ref{H_Zag}) for $k=2$) the
evolution operator (\ref{1}) takes the canonical form of the quantum
BBGKY hierarchy \cite{BogLect}
\begin{eqnarray*}
      &&i\hbar\frac{\partial}{\partial t} F_{s}(t;q_{1},\ldots,q_{s};q^{'}_{1},\ldots,q^{'}_{s})=
      \Big(-\frac{\hbar^{2}}{2}\sum\limits_{i=1}^{s}(\Delta_{q_{i}}-\Delta_{q^{'}_{i}})
      +\nonumber\\
      &&\qquad+\sum\limits_{i<j=1}^{s}\big(\Phi(q_{i}-q_{j})-\Phi(q^{'}_{i}-q^{'}_{j})\big)\Big)
      F_{s}(t;q_{1},\ldots,q_{s};q^{'}_{1},\ldots,q^{'}_{s})+ \\
      &&\qquad+\sum\limits_{i=1}^{s}\int dq_{s+1}\big(
      \Phi(q_{i}-q_{s+1})-\Phi(q^{'}_{i}-q_{s+1})\big)F_{s+1}(t;q_{1},\ldots
      q_{s},q_{s+1};q^{'}_{1},\ldots,q^{'}_{s},q_{s+1}).\nonumber
\end{eqnarray*}
For the solution of the BBGKY hierarchy the following theorem is true \cite{GerS}.
\begin{theorem} If $F(0)\in\mathfrak{L}_{\alpha}^{1}(\mathcal{F} _\mathcal{H})$ and $\alpha>e$, then for $t\in\mathbb{R}^{1}$ there
exists a unique solution of initial-value problem (\ref{1})-(\ref{2}) given by
\begin{multline}\label{RozvBBGKY}
      F_{s}(t,1,\ldots,s)=\\=\sum\limits_{n=0}^{\infty}\frac{1}{n!}\mathrm{Tr}_{s+1,\ldots,{s+n}}
      \mathfrak{A}_{1+n}(t,Y_{1},s+1,\ldots,s+n)F_{s+n}(0,1,\ldots,s+n),\quad s\geq1,
\end{multline}
where for $n \geq 0$
\begin{eqnarray*}
\mathfrak{A}_{1+n}(t,Y_{1},s+1,\ldots,s+n)=\sum\limits_{\mathrm{P}:\{Y_{1},X\setminus Y\} =\bigcup\limits_iX_i}(-1)^{|\mathrm{P}|-1}(|\mathrm{P}|-1)!
        \prod_{X_i\subset \mathrm{P}}\mathcal{G}_{|X_i|}(-t,X_i)
\end{eqnarray*} is the $(1+n)th$-order cumulant  of operators (\ref{groupG}),
$X\setminus Y=\{s+1,\ldots,s+n\}$ and $Y_{1}=\{1\cup\ldots\cup s\}$.

For initial data $F(0)\in\mathfrak{L}^{1}_{\alpha,0}\subset
\mathfrak{L}^{1}_{\alpha}(\mathcal{F}_\mathcal{H})$ it is a strong
solution and for arbitrary initial data from the space
$\mathfrak{L}_{\alpha}^{1}(\mathcal{F}_\mathcal{H})$ it is a weak
solution.
\end{theorem}
The condition $ \alpha> e $ guarantees the convergence of series
(\ref{RozvBBGKY}) and means that the average number
 of particles (\ref{averageND}) is finite: $\langle
N\rangle<\alpha/e.$ This fact follows if to renormalize a sequence
(\ref{RozvBBGKY}) in such a way:
 $\widetilde{F}_{s}(t)=\langle N\rangle^{s}F_{s}(t)$.

For arbitrary
$F(0)\in\mathfrak{L}^{1}_{\alpha}(\mathcal{F}_\mathcal{H})$  the
average number  of particles (expectation value (\ref{averageND}) expressed in terms of $s$-particle operators)
\begin{equation}\label{N_F}
\langle N\rangle(t) = \mathrm{Tr}_{1} F_{1}(t,1)
\end{equation}
 in state (\ref{RozvBBGKY}) is
finite, in fact,
\begin{equation*}
| \langle N\rangle(t)| \leq
c_{\alpha}\|F(0)\|_{\mathfrak{L}_{\alpha}^{1}(\mathcal{F}_\mathcal{H})}<\infty,
\end{equation*}
where $c_{\alpha}=e^{2}(1-\frac{e}{\alpha})^{-1}$ is a constant. We
emphasize the difference between finite and infinite systems with
non-fixed number of particles. To describe an infinite particle
system we have to construct a solution of initial-value problem
(\ref{1})-(\ref{2}) in more general spaces than
$\mathfrak{L}_{\alpha}^{1}(\mathcal{F}_\mathcal{H}),$ for example,
in the space of sequences of bounded operators to which the
equilibrium states belong \cite{Rul,Gen}.

We remark that the formula for solution (\ref{RozvBBGKY}) can be
directly derived from solution (\ref{rozvNF-N}) of von Neumann
hierarchy (\ref{nelNeum1}) \cite{Dop08}. In papers
\cite{AA,Gol,Petr72,Schlein} a solution of initial-value problem
(\ref{1})-(\ref{2}) is represented as the perturbation (iteration)
series, which for a two-body interaction potential has the form
            \begin{multline}\label{Iter}
                 F_{s}(t)=\sum\limits_{n=0}^{\infty}\int\limits_0^tdt_{1}\ldots\int\limits_0^{t_{n-1}}dt_{n}\mathrm{Tr}_{s+1,\ldots,s+n}
                 \mathcal{G}_{s}(-t+t_{1})\sum\limits_{i_{1}=1}^{s}\big(-\mathcal{N}^{(2)}_{int}(i_{1},s+1)\big)
                    \mathcal{G}_{s+1}(-t_{1}+t_{2})\times\\
            \ldots \mathcal{G}_{s+n-1}(-t_{n-1}+t_{n})
                  \sum\limits_{i_{n}=1}^{s+n-1}\big(-\mathcal{N}^{(2)}_{int}(i_{n},s+n)\big)
                      \mathcal{G}_{s+n}(-t_{n})F_{s+n}(0).
                \end{multline}
In the space $\mathfrak{L}_{\alpha}^{1}(\mathcal{F}_\mathcal{H})$
expansion (\ref{1}) is equivalent to this iteration series. This
follows from the validity of analogs of the Duhamel formula for
cumulants of evolution operators of the von Neumann equations. For
the second-order cumulant an analog of the Duhamel formula has the
form (\ref{iter2kum}) and in the general case the following formula
takes place
\begin{multline*}\label{KymN}
     \mathfrak{A}_{1+n}(t,Y_{1},s+1,\ldots,s+n)=
     \\=\int\limits_0^tdt_{1}\ldots\int\limits_0^{t_{n-1}}dt_{n}
     \prod\limits_{k_{1}\in\mathcal{I}_{1}}\mathcal{G}_{|k_{1}|}(-t+t_{1},k_{1})
      \sum\limits_{i_{1}<j_{1}\in \mathcal{I}_{1}}
        \sum\limits_{l_{1}\in i_{1}}
        \sum\limits_{m_{1}\in j_{1}}
                        \big(-\mathcal{N}^{(2)}_{int}(l_{1},m_{1})\big)
       \times\\
\times\prod\limits_{k_{2}\in\mathcal{I}_{2}}\mathcal{G}_{|k_{2}|}(-t_{1}+t_{2},k_{2})
      \sum\limits_{i_{2}<j_{2}\in \mathcal{I}_{2}}
                \sum\limits_{l_{2}\in i_{2}}
        \sum\limits_{m_{2}\in j_{2}}
         \big(-\mathcal{N}^{(2)}_{int}(l_{2},m_{2})\big)
       \times\ldots\\\ldots
      \times \prod\limits_{k_{n}\in\mathcal{I}_{n}}\mathcal{G}_{|k_{n}|}(-t_{n-1}+t_{n},k_{n})
      \sum\limits_{i_{{n}}<j_{{n}}\in \mathcal{I}_{n}}
                \sum\limits_{l_{n}\in i_{n}}
        \sum\limits_{m_{n}\in j_{n}}
        \big(-\mathcal{N}^{(2)}_{int}(l_{n},m_{n})\big)\mathcal{G}_{s+n}(-t_{n},1,\ldots,s+n),
    \end{multline*}
 where $\mathcal{I}_{1}\equiv \{Y_{1},s+1,\ldots,s+n\}$,
$\mathcal{I}_{n}\equiv\{i_{{n-1}}\cup j_{{n-1}}\}\cup
\mathcal{I}_{n-1}\setminus\{i_{{n-1}},i_{{n-1}}\}$.

We remark that  for classical systems of particles the first few
terms of the cumulant expansion (\ref{RozvBBGKY}) were considered in
 \cite{Co68}.

In \cite{GerS} we discuss other possible representations of a
solution of the BBGKY hierarchy in the space
$\mathfrak{L}_{\alpha}^{1}(\mathcal{F}_\mathcal{H}).$

\subsection{Correlation operators of infinite-particle systems}

Correlation operators (\ref{rozvNF-N}) may be employed to directly calculate the macroscopic values
 of a system, in particular, fluctuations characterized by
the average values of the square deviations of observables from its
average values. For example, for an additive-type observable
$a=(a_{0},a_{1}(1),\ldots,\sum\limits_{i=1}^{n}a_{1}(i),\ldots)$
from the formula for expectation value (\ref{averageD}) we derive
the formula for fluctuations (the dispersion of an additive-type
observable)  \cite{BogLect}
\begin{multline*}
\langle\big(a-\langle
a\rangle(t)\big)^{2}\rangle(t)=\\=\mathrm{Tr}_{1}\big(a_{1}^{2}(1)-\langle
a\rangle^{2}(t)\big)F_{1}(t,1)+\mathrm{Tr}_{1,2}a_{1}(1)a_{1}(2)\big(F_{2}(t,1,2)-F_{1}(t,1)F_{1}(t,2)\big).
\end{multline*}
Therefore, the dispersion of the additive-type observable is defined
not directly through the solutions of the BBGKY hierarchy but by the
following correlation operators:\linebreak
$F_{2}(t,1,2)-F_{1}(t,1)F_{1}(t,2)=G_{2}(t,1,2)$ or in the general
case by the \emph{$s$-particle correlation operators}
\begin{eqnarray}\label{G(F)}
       G_{s}(t,1,...,s):=\sum\limits_{\texttt{P}:\,\{1,...,s\}=\bigcup\limits_iX_i}(-1)^{\abs {\texttt{P}}-1}(\abs {\texttt{P}} -1)!\,
        \prod_{X_i\subset \texttt{P}}F_{\abs {X_i}}(t,X_i).
\end{eqnarray}
 The $s$-particle correlation operators $G_{s}(t),$ $s\geq 1$ can be expressed in terms of correlation operators (\ref{rozvNF-N}) by the formula
\begin{eqnarray}\label{G(g)comp}
G_{s}(t,1,...,s)=\sum\limits_{n=0}^{\infty}\frac{1}{n!}\mathrm{Tr}_{s+1,\ldots,s+n}g_{s+n}(t,1,\ldots,s+n),\quad\quad s\geq1,
\end{eqnarray}
where $g_{s+n}(t,1,\ldots,s+n)$ is a solution of the von Neumann hierarchy (\ref{nelNeum1}).

To derive expression (\ref{G(g)comp}) we state the following lemma.

\begin{lemma}
 Let $f=(f_{0},f_{1},\ldots,f_{n})$ and $f_{n}\in\mathfrak{L}^{1}(\mathcal{H}_{n}),$ then the equality holds
      \begin{eqnarray*}
      (e^\mathfrak{a}\mathbb{E}\mathrm{xp}_{\ast} f)^{-1}_{0}e^\mathfrak{a}\mathbb{E}\mathrm{xp}_{\ast} f
      =\mathbb{E}\mathrm{xp}_{\ast}
      e^\mathfrak{a}f.
      \end{eqnarray*}
\end{lemma}
     \begin{proof}
      Indeed, using equality (\ref{efg}) and the equality
      \begin{equation*}
      \mathfrak{d}_{1}\ldots\mathfrak{d}_{s}\mathbb{E}\mathrm{xp}_{\ast}f=\mathbb{E}\mathrm{xp}_{\ast} f\ast\sum\limits_{\texttt{P}:Y=\bigcup_i X_{i}}
      \mathfrak{d}_{1}\ldots\mathfrak{d}_{|X_1|}f\ast\ldots\ast \mathfrak{d}_{|X_{|P|-1}|+1}
      \ldots\mathfrak{d}_{|X_{|\texttt{P}|}|}f,
      \end{equation*} that follows from (\ref{oper_d}), (\ref{d_gamma}),
      one obtains (here  $Y=(1,\ldots,s)$)
      \begin{multline*}
      (e^\mathfrak{a}\mathbb{E}\mathrm{xp}_{\ast} f)^{-1}_{0}(e^\mathfrak{a}\mathbb{E}\mathrm{xp}_{\ast} f)_{|Y|}(Y)
      =(e^\mathfrak{a}\mathbb{E}\mathrm{xp}_{\ast} f)^{-1}_{0}(e^\mathfrak{a}\mathfrak{d}_{1}\ldots\mathfrak{d}_{s}(\mathbb{E}\mathrm{xp}_{\ast} f))_{0}=\\
      =(e^\mathfrak{a}\mathbb{E}\mathrm{xp}_{\ast} f)^{-1}_{0}\bigg(e^\mathfrak{a}(\mathbb{E}\mathrm{xp}_{\ast} f\ast
      \sum\limits_{\texttt{P}:Y=\bigcup X_{i}}\mathfrak{d}_{1}\ldots\mathfrak{d}_{|X_1|}f\ast
      \ldots
\mathfrak{d}_{|X_{|P|-1}|+1}
      \ldots\mathfrak{d}_{|X_{|\texttt{P}|}|}f)\bigg)_{0}=\nonumber\\
      =(e^\mathfrak{a}\mathbb{E}\mathrm{xp}_{\ast} f)^{-1}_{0}(e^\mathfrak{a}\mathbb{E}\mathrm{xp}_{\ast} f)_{0}
      \bigg(e^\mathfrak{a}\sum\limits_{\texttt{P}:Y=\bigcup
      X_{i}}\mathfrak{d}_{1}\ldots\mathfrak{d}_{|X_1|}f\ast\ldots
      \ast \mathfrak{d}_{|X_{|\texttt{P}|-1}|+1}
      \ldots\mathfrak{d}_{|X_{|\texttt{P}|}|}f\bigg)_{0}=\nonumber\\
      =\sum\limits_{\texttt{P}:\,Y=\bigcup\limits_iX_i}\,
      \prod_{X_i\subset
      \texttt{P}}(e^\mathfrak{a}\mathfrak{d}_{1}\ldots\mathfrak{d}_{|X_{i}|}f)_{0}=\big(\mathbb{E}\mathrm{xp}_{\ast}
      e^\mathfrak{a}f\big)_{|Y|}(Y).\nonumber
      \end{multline*}
      \end{proof}
We now derive representation (\ref{G(g)comp}) for $s$-particle correlation operators. Using equality
(\ref{D_(g)N}) in terms of the mapping
$\mathbb{E}\mathrm{xp}_{\ast}$ (\ref{Exp}), representation (\ref{F(D)}) for $s$-particle statistical operators can be rewritten in the form
$$F(t)=(e^\mathfrak{a}\mathbb{E}\mathrm{xp}_{\ast}g(0))^{-1}_{0}e^{\mathfrak{a}}\mathbb{E}\mathrm{xp}_{\ast}
g(t).$$
In terms of the mapping $\mathbb{E}\mathrm{xp}_{\ast}$
formula (\ref{G(F)})
 has the form
     \begin{equation*}\label{F(G)}
      F(t)=\mathbb{E}\mathrm{xp}_{\ast} G(t).
     \end{equation*}
Further, according to previous formula and Lemma 3  we have
     \begin{equation*}
      \mathbb{E}\mathrm{xp}_{\ast} G(t)=(e^\mathfrak{a}\mathbb{E}\mathrm{xp}_{\ast} g)^{-1}_{0}e^{\mathfrak{a}}\mathbb{E}\mathrm{xp}_{\ast} g(t)=\mathbb{E}\mathrm{xp}_{\ast}
     e^{\mathfrak{a}}g(t),
     \end{equation*}
and, therefore, we finally derive
     \begin{equation}\label{G(g)}
     G(t)=
     e^{\mathfrak{a}}g(t)
     \end{equation}
or in component-wise form (\ref{G(g)comp}).

 For chaos initial data (\ref{posl_g(0)}) obeying Maxwell-Boltzmann statistics according to (\ref{G(g)}) we have
$G_{1}(0)=g_{1}(0).$ Using  solution (\ref{rozvChaosN}) of the von
Neumann hierarchy (\ref{nelNeum1}) the expansion for a solution  of
the initial-value problem for $s$-particle correlation operators can
be represented as follows
\begin{equation*}\label{GUg}
     G_{s}(t,1,\ldots,s)
     =\sum\limits_{n=0}^{\infty}\frac{1}{n!}\mathrm{Tr}
      _{{s+1,}\ldots ,s+n}\mathfrak{A}_{s+n}(t,1,\ldots,s+n)
        \prod_{i=1}^{s+n}G_{1}(0,i), \quad s\geq 1.
\end{equation*}
It should be noted that  sequence (\ref{G(g)}) is a solution of the nonlinear BBGKY hierarchy
for $s$-particle correlation operators \cite{Gre56} which  describes the dynamics of correlations  of
      infinite-particle systems.

\section*{Acknowledgement}
This work was partially supported by the WTZ grant No M/124 (UA 04/2007) and by the Special programm of the PAD of NAS of Ukraine.

\end{document}